\documentclass[aps,prd,groupedaddress,preprint,eqsecnum,nofootinbib,showpacs]{revtex4}
\usepackage{graphicx,epsf}
\newif\ifdraft
\drafttrue
\newif\ifpreprint
\preprinttrue

\def\sect#1{section~{\ref{#1}}}
\def\fig#1{fig.~{\ref{#1}}}

\def\tab#1{table~{\ref{#1}}}

\def\spa#1.#2{\left\langle#1\,#2\right\rangle}
\def\spb#1.#2{\left[#1\,#2\right]}
\def\spash#1.#2{\spa{\smash{#1}}.{\smash{#2}}}
\def\spbsh#1.#2{\spb{\smash{#1}}.{\smash{#2}}}
\def\sand#1.#2.#3{%
\left\langle\smash{#1}{\vphantom1}^{-}\right|{#2}%
\left|\smash{#3}{\vphantom1}^{-}\right\rangle}
\def\sandpp#1.#2.#3{%
\left\langle\smash{#1}{\vphantom1}^{+}\right|{#2}%
\left|\smash{#3}{\vphantom1}^{+}\right\rangle}
\def\sandpm#1.#2.#3{%
\left\langle\smash{#1}{\vphantom1}^{+}\right|{#2}%
\left|\smash{#3}{\vphantom1}^{-}\right\rangle}
\def\sandmp#1.#2.#3{%
\left\langle\smash{#1}{\vphantom1}^{-}\right|{#2}%
\left|\smash{#3}{\vphantom1}^{+}\right\rangle}
\def\Shift#1#2{{[#1,#2\rangle}}

\def\tree{{\rm tree}}
\def\pol{\varepsilon}
\def\Tr{\, {\rm Tr}}

\def\eps{\epsilon}
\def\e{\epsilon}

\def\nn{\nonumber}

\def\eqn#1{eq.~(\ref{#1})}

\def\eqns#1#2{eqs.~(\ref{#1}) and~(\ref{#2})}

\def\NeqFour{{{\cal N}=4}}

\def\NeqEight{{{\cal N}=8}}
\def\NeqOne{{{\cal N}=1}}

\def\be{\begin{equation}}
\def\ee{\end{equation}}
\def\bea{\begin{eqnarray}}
\def\eea{\end{eqnarray}}
\def\ba{\begin{eqnarray}}
\def\ea{\end{eqnarray}}

\def\ksl{\s{k}}
\def\Perm{{\cal P}}
\def\ve{\varepsilon}
\def\tlambda{{\tilde\lambda}}
\def\MHVbar{$\overline{\hbox{MHV}}$}

\newcommand{\Bmp}[1]{\langle #1\rangle}

\newcommand{\Kf}[1]{K^{\flat}_{#1}}

\newcommand{\Kfm}[1]{K^{\flat,-}_{#1}}

\def\tree{{\rm tree}}
\def\oneloop{{\rm 1\hbox{-}loop}}

\def\Ord{{\cal O}}

\newbox\charbox
\newbox\slabox
\def\s#1{{      % Feynman slash
        \setbox\charbox=\hbox{$#1$}
        \setbox\slabox=\hbox{$/$}
        \dimen\charbox=\ht\slabox
        \advance\dimen\charbox by -\dp\slabox
        \advance\dimen\charbox by -\ht\charbox
        \advance\dimen\charbox by \dp\charbox
        \divide\dimen\charbox by 2
        \raise-\dimen\charbox\hbox to \wd\charbox{\hss/\hss}
        \llap{$#1$} }}
\def\ksl{\s{k}}

\begin{document}

\ifpreprint
UCLA/07/TEP/16
\hfill $\null\hskip 1.0 cm \null$  SLAC--PUB--12609
\fi

\title{Unexpected Cancellations in Gravity Theories}

\iffalse
\author{Zvi Bern}
\affiliation{{} Department of Physics and Astronomy, UCLA\\
\hbox{Los Angeles, CA 90095--1547, USA} }

\author{Darren Forde}
\affiliation{ Department of Physics and Astronomy, UCLA\\
\hbox{Los Angeles, CA 90095--1547, USA} \\
and \\
Stanford Linear Accelerator Center \\
             Stanford University\\
             Stanford, CA 94309, USA}

\author{Harald Ita}
\affiliation{Department of Physics\\
Swansea University \\
Swansea, SA2 8PP, UK}
\fi

\author{Z.~Bern${}^a$,  J.~J.~Carrasco${}^a$, D.~Forde${}^{a,b}$,
H. Ita${}^{c}$ and H.~Johansson${}^a$}

\affiliation{
${}^a$Department of Physics and Astronomy, UCLA, Los Angeles, CA
90095-1547, USA  \\
${}^b$Stanford Linear Accelerator Center,
              Stanford University,
             Stanford, CA 94309, USA \\
${}^c$Department of Physics, Swansea University, Swansea, SA2 8PP, UK}

\date{July, 2007}

\begin{abstract}
Recent computations of scattering amplitudes show that  
$\NeqEight$ supergravity is surprisingly well behaved
in the ultraviolet and may even be ultraviolet finite in perturbation
theory.  The novel cancellations necessary for ultraviolet finiteness
first appear at one loop in the guise of the ``no-triangle
hypothesis''.  We study one-loop amplitudes in pure Einstein gravity
and point out the existence of cancellations similar to those
found previously in $\NeqEight$ supergravity.  These cancellations go
beyond those found in the one-loop effective action. Using unitarity,
this suggests that generic theories of quantum gravity based on the
Einstein-Hilbert action may be better behaved in the ultraviolet at
higher loops than suggested by naive power counting, though without
additional (supersymmetric) cancellations they diverge.  We comment
on future studies that should be performed to support this proposal.
\end{abstract}

\pacs{11.15.Bt, 11.25.Db, 11.25.Tq, 11.55.Bq, 12.38.Bx \hspace{1cm}}

\maketitle

%%%%%%%%%%%%%%%%%%%%%%%%%%%%%%%%

\section{Introduction}

Recent calculations of four-point scattering amplitudes in the
maximally supersymmetric gravity theory of Cremmer and
Julia~\cite{CremmerJuliaScherk} indicate the existence of novel
ultraviolet cancellations at three and higher loops, which may even
lead to the perturbative finiteness of the
theory~\cite{Finite,GravityThree}.  String dualities have also been
used to argue for ultraviolet finiteness of $\NeqEight$
supergravity~\cite{DualityArguments}, though difficulties with
decoupling towers of massive states may alter this
conclusion~\cite{GSO}.

At one loop in the $\NeqEight$ theory, the corresponding novel
cancellations are encapsulated in the ``no-triangle
hypothesis''~\cite{OneloopMHVGravity,NoTriangle,SixPointIR,NoTriangleSixPt}. In
general, dimensionally regularized amplitudes in four dimensions can
be expressed as a linear combination of scalar box, triangle and
bubble integrals together with rational terms~\cite{PV,DimRegPV}.  The
no-triangle hypothesis states that all one-loop amplitudes in
$\NeqEight$ supergravity can be expressed solely in terms of scalar
box integrals and that neither triangle integrals, bubble integrals
nor additional rational terms appear. This hypothesis is surprising
from the point of view of power counting, which --- together with
standard integral reduction formulas --- would imply that triangle
integrals should appear beginning at five points and bubble integrals
at six points.  The cancellation of triangle and bubble integrals were
first observed in maximally helicity violating (MHV) amplitudes of the
$\NeqEight$ theory~\cite{OneloopMHVGravity}. More recently, this
observation was extended to the hypothesis that the same cancellations
occur in {\it all} $\NeqEight$ one-loop amplitudes, so that they are
expressed solely in terms of scalar box
integrals~\cite{NoTriangle}. It has been confirmed for six-graviton
amplitudes, by explicit computation~\cite{NoTriangleSixPt}.
Furthermore, at seven points the infrared singularities have been
shown to be consistent with the no-triangle
hypothesis~\cite{SixPointIR,NoTriangleSixPt}.  Beyond six and seven
points, scaling and factorization properties of the amplitudes provide
strong evidence that the no-triangle hypothesis holds for the
remaining amplitudes in $\NeqEight$
supergravity~\cite{NoTriangle,NoTriangleSixPt}.  In
ref.~\cite{Finite}, these one-loop cancellations were argued to lead
to an improved ultraviolet behavior in classes of terms encountered
in multi-loop calculations in the $\NeqEight$ theory.  At three loops,
the improvement in the ultraviolet behavior due to these and related
cancellations has been confirmed by the explicit calculation of the
complete four-point scattering amplitude in the $\NeqEight$
theory~\cite{GravityThree}.  The consistency of the Regge limit with
improved ultraviolet behavior in $\NeqEight$ supergravity has also
been recently discussed in ref.~\cite{Schnitzer}.

Supersymmetry, in particular, has long been studied for its ability to
reduce the degree of divergence of gravity
theories~\cite{SuperGravity}.  However, all superspace power counting
arguments ultimately delay the onset of divergences only by a finite
number of loops, depending upon assumptions regarding the types of
superspaces and invariants that can be constructed.

If the cancellations observed in the $\NeqEight$ theory are only
partly due to supersymmetry, then what might account for the remainder
of the observed cancellations?  Here we propose that {\it these extra
cancellations are generic to any quantum gravity theory} based on the
Einstein-Hilbert action.  These cancellations are not at all obvious in
Feynman diagrams which individually obey a naive power
counting.  Rather, they will be manifest only in carefully chosen
representations of the amplitudes.  Our proposal is that in
supersymmetric theories, the supersymmetric cancellations are on top
of these cancellations.  
 For the $\NeqEight$ theory at one loop it is 
the combination of cancellations that leads to the ``no-triangle
hypothesis''.  Following the same line of reasoning as used in 
the $\NeqEight$ theory~\cite{Finite}, suggests
that the observed one-loop cancellations in non-supersymmetric pure gravity
will lead to a softening of the ultraviolet behavior at higher loops.

For non-supersymmetric theories these additional cancellations are in
general insufficient to render the theory ultraviolet finite.  Indeed,
explicit calculations show that gravity coupled to matter generically
diverges at one loop~\cite{tHooftVeltmanGravity}.  Pure Einstein
gravity in four dimensions possesses a cancellation at one loop,
distinct from the ones being discussed here, due to
the absence of a viable counterterm, delaying the divergence to two
loops.  This two-loop divergence was established by Goroff and
Sagnotti through direct computation of that
divergence~\cite{GoroffSagnotti} and later confirmed by van de
Ven~\cite{vandeVen}.

Here, as a first step in checking the hypothesis that there 
are novel cancellations in non-supersymmetric gravity theories,
we will investigate pure gravity at one loop.  To carry
out our investigation of cancellations we make use of the unitarity
method~\cite{UnitarityMethod, Fusing,BDDPR}.  In order to observe the
cancellations we reduce the amplitudes to combinations of box, bubble
and triangle integrals. The cancellations then manifest themselves in
unexpectedly low powers of loop momentum in these integrals. The
reduction to the basic integrals allows us to combine the
contributions coming from the higher-point integrals to make the
cancellations explicit.  We do so using powerful new loop integration
methods~\cite{BCFUnitarity, BBCFOneLoop, OPP, ZoltanDdim,
NoTriangleSixPt, Forde}, based on generalized
unitarity~\cite{GeneralizedUnitarity} and complex
momenta~\cite{GoroffSagnotti,WittenTopologicalString}.  In particular, we make
extensive use of the formalism introduced in ref.~\cite{Forde}. This
will allow us to observe the hidden cancellations through simple scaling
arguments.

The unitarity method requires as input tree amplitudes, which we
construct via Kawai-Lewellen-Tye (KLT) relations~\cite{KLT,GeneralKLT} and
on-shell recursion relations~\cite{BCFRecursion,BCFW, GravityRecursion,
CachazoLargez}.
For MHV tree amplitudes, we use the form obtained by Berends, Giele
and Kuijf \cite{BGK} via the KLT relations.  This is a particularly
compact form, making apparent various useful properties of the
amplitudes.

The one-loop cancellations we observe rely on rather generic
properties of gravity tree-level amplitudes. In particular, certain
scaling properties of the tree amplitudes have to be independent of
the number of scattering particles.  These cancellations are quite
reminiscent of ones that occur at tree level under the large complex
deformations needed to prove the validity of on-shell recursion
relations~\cite{GravityRecursion, CachazoLargez}.
Indeed, the connection of large complex deformations to the absence of
scalar bubble integrals in $\NeqEight$ supergravity has been already
noted in ref.~\cite{NoTriangleSixPt}. This connection suggests that
the cancellations will be present in theories that are ``fully
constructible'' from on-shell recursion relations, in the sense of
ref.~\cite{CachazoConstuctible}.  In such theories, the tree amplitudes
can be obtained using on-shell recursion using only three vertices 
as the input.  Remarkably, Einstein gravity is in this class.

This paper is organized as follows.  In \sect{TreeSection}, we review
properties of gravity tree amplitudes, including the cancellations
that are present under large complex deformations.  Then in
\sect{OneLoopSection} we describe the methods we use to count the
leading powers of loop momentum in the triangle and bubble integrals
contributing in Einstein gravity.  In this section we also contrast
these cancellations against the well known one which renders pure
Einstein gravity finite at one loop.  We present the explicit power
counting of the triangle and bubble integrals in
\sect{PowerCountingSection}.  In \sect{CancellationSection} we explain
the relationship between the cancellations we observe at one loop and
ones that were observed previously at tree level. We also give a
heuristic explanation of observed one-loop cancellations, based on
universal factorization properties of one-loop amplitudes.  We
conclude and comment on the outlook in \sect{ConclusionSection}.

\section{Properties of tree amplitudes}
\label{TreeSection}

In this section, we first review some well known representations of
tree amplitudes, used later for explicit computations, and then review
cancellations in tree amplitudes needed for on-shell recursion relations.

\subsection{Notation for gravity tree amplitudes}
 
To expose useful properties of scattering amplitudes in four dimensions
we  employ the spinor-helicity
formalism~\cite{SpinorHelicity,TreeReview}, with spinor products,
\begin{eqnarray}
\spa{j}.{l}
&=& \ve^{\alpha\beta} \lambda_{j\alpha} \lambda_{l\beta}
= \bar{u}_-(k_j) u_+(k_l)
\label{spinorproddefa}\,,\quad\quad 
\spb{j}.{l}
= \ve^{\dot\alpha\dot\beta} 
\tlambda_{j\dot\alpha} \tlambda_{l\dot\beta}
= \bar{u}_+(k_j) u_-(k_l)\,,
\label{spinorproddefb}
 \end{eqnarray}  
where $u_{\pm}(k)$ is a massless Weyl spinor with momentum $k$ and
positive or negative chirality. It will also be convenient to use the
bra-ket notation for the contractions of Weyl spinors; $ \langle j^\mp
| l^\pm \rangle =\bar{u}_\mp(k_j) u_\pm(k_l)$. With the normalizations
used here, the spinor inner products are related to Lorentz inner products via,
$\spa{l}.{j} \spb{j}.{l} = {1\over2} \Tr[ \ksl_j \ksl_l ] = 2k_j\cdot
k_l = s_{jl}$.
A useful identity is the Schouten identity,
\begin{equation}
\spa{1}.{2} \spa{3}.{4} = 
  \spa{2}.{3} \spa{4}.{1} + \spa{2}.{4} \spa{1}.{3} \,,
\label{Schouten}
\end{equation} 
valid for four arbitrary spinors $\lambda_1,\lambda_2,\lambda_3$ and
$\lambda_4$.
In the spinor-helicity formalism 
gluon (spin $1$) polarization vectors take the form,
\begin{equation}
\pol^\pm_{\mu}(k_i, q_i)  = 
\frac{1}{2}\, {\langle q_i^\mp | \gamma_\mu| k_i^\mp \rangle \over
 \langle q_i^\mp | k_i^\pm \rangle }\,,
\label{GluonPol}
\end{equation} 
where $k_i$ is the momentum carried by the particle and $q_i$ is a
null reference momentum.  
The graviton polarization is,
\begin{equation}
\pol^\pm_{\mu\nu}(k_i, q_i) =
 \pol^\pm_{\mu}(k_i, q_i)\pol^\pm_{\nu}(k_i, q_i)\, .
\label{GravitonPol}
\end{equation}
It is worth noting that the tracelessness condition $\pol_\mu{}^\mu = 0$
is automatically enforced since $\langle q_i^\mp | \gamma_\mu| k_i^\mp
\rangle$ is a complex null vector.

A particularly useful representation for gravity tree amplitudes is
based on the Kawai, Lewellen and Tye relations~\cite{KLT} between open
and closed string theory tree-level amplitudes, especially since this
exposes the intimate connection between gauge and gravity
amplitudes. In the low-energy limit these become relations between
gravity and gauge theory amplitudes.  Not only do they hold for any
tree amplitudes obtained from the low-energy limit of a string theory,
but they appear to hold more generally~\cite{GeneralKLT}.  For three
through six points the relations are,
\begin{eqnarray}  
M_3^{\tree}(1,2,3) & = &i A_3^{\rm tree} (1,2,3) \, A_4^{\rm tree}(1,3,2)\,,
\label{KLTThreePoint} \\
M_4^{\tree}  (1,2,3,4) &=& 
 - i s_{12} A_4^{\rm tree} (1,2,3,4) \, A_4^{\rm tree}(1,2,4,3)\,,
\label{KLTFourPoint} \\
M_5^{\rm tree}(1,2,3,4,5) & = &
i s_{12} s_{34}  A_5^{\rm tree}(1,2,3,4,5)
                                 A_5^{\rm tree}(2,1,4,3,5) \nonumber\\
&& \hskip .3 cm \null
 + i s_{13}s_{24} A_5^{\rm tree}(1,3,2,4,5) \, A_5^{\rm tree}(3,1,4,2,5)\,,
\label{KLTFivePoint} \\
M_6^\tree(1,2,3,4,5,6) 
& = & - i s_{12} s_{45}  A_6^\tree(1,2,3,4,5,6) 
    \Bigl[ s_{35} A_6^\tree(2,1,5,3,4,6)  \nn\\
&& \hskip 1 cm 
   + (s_{34} + s_{35}) A_6^\tree(2,1,5,4,3,6) \Bigr] 
\ +\ \Perm(2,3,4) \,. 
\label{KLTSixPoint}
\end{eqnarray}
Expressions for any numbers of legs may be found in appendix A of
ref.~\cite{OneloopMHVGravity}.  Here the $M_n^\tree$'s are $n$-point
tree-level amplitudes in a gravity theory.  The $A_n^\tree$'s are
color-ordered gauge-theory partial amplitudes, with $A_n^\tree(1, 2,
\ldots n)$ representing the kinematic coefficient of the color trace,
$\Tr[T^{a_1} T^{a_2} \cdots T^{a_n}]$.  In the six-point case, in
\eqn{KLTSixPoint}, $\Perm(2,3,4)$ signifies a sum over permutations
over the legs $2,3$ and $4$.  We have suppressed factors of the
coupling constants in the gauge theory and gravity amplitudes.  The
KLT relations are valid for any configuration of helicities and
particles. As
an abbreviation we use the labels ``$1,\ldots, n$'' to denote the
momenta $k_1,\ldots, k_n$ and polarizations or spinors of external legs.

For example, using known expressions for the gauge theory
amplitudes~\cite{TreeReview,OnShellReview}, for $n=3,4$ the KLT
relations give us,
\begin{eqnarray}
M_3^\tree(1^-,2^-,3^+) &=& i {\spa{1}.{2}^6 \over \spa{2}.{3}^2
 \spa{3}.{1}^2} \,, \nn\\
M_4^\tree(1^-,2^-, 3^+, 4^+) &=& i {\spb{3}.{4} \spa{1}.{2}^6 \over 
                \spa{2}.{3} \spa{3}.{4} \spa{4}.{1} \spa{2}.{4} \spa{3}.{1}}\,.
\label{ThreeFourPointTree}
\end{eqnarray}

Although very useful, the KLT form often makes all-$n$ analyses
difficult due to large permutation sums and non-manifest factorization
and scaling properties. For the specific case of MHV all-$n$ ($n>4$) graviton
amplitudes, Berends, Giele and Kuijf (BGK)~\cite{BGK} presented the
more manageable expression,
\begin{eqnarray}
&&\hspace*{-1.cm}M_n^\tree(1^+,2^+,\ldots, a^-, \ldots, b^-,\ldots,n^+)=
\label{BGKMHV}
\nn\\
&&\hspace*{-0.5cm}
-i \spa{a}.{b}^8\sum_{\Perm(3,4,\ldots,n-1)}
{\prod_{l=3}^{n-1} \sand{n}.{\s K_{2\ldots(l-1)}}.l \over
 \prod_{i=1}^{n-2}\spa{i}.{(i+1)}\spa{1}.{(n-1)}\spa1.n^2
  \spa2.n^2 \prod_{l=3}^{n-1}\spa{l}.{n}}\,,
\end{eqnarray}
where all legs besides $a$ and $b$ are of positive helicity, and
$K^\mu_{2 \ldots (l-1)} \equiv \sum_{i =2}^{l-1} k_i^\mu$, and
$\Perm(3, 4, \ldots, n-1)$ indicates the summation over all
permutations of the labels $3, 4, \ldots,n-1$. (The form here is
slightly rearranged compared to the one in ref.~\cite{BGK}.) 
This expression has been checked
numerically through $11$-points against the Kawai, Lewellen and Tye
relations~\cite{BGK}, as well as those derived from on-shell
recursion~\cite{GravityRecursion}.  Although the BGK formula has been
confirmed only through 11 points, the fact that it has the correct
properties as any momentum becomes soft, makes it extremely likely to
be correct to all $n$.  Although not manifestly so, 
this formula is fully crossing symmetric under an
interchange of any pairs of legs, after dividing by
$\spa{a}.{b}^8$.  The \MHVbar\ graviton amplitudes,
which have two positive helicity graviton legs and the rest negative,
are obtained simply by swapping angle with square brackets.

The MHV tree amplitudes satisfy simple supersymmetry Ward
identities~\cite{SWI}, allowing us to replace two of the graviton legs
with any pair of particles of lesser spin,
\begin{eqnarray}
M_n^\tree(1^{-h}, 2^-, 3^+, \ldots, n^h) = 
\biggl({\spa2.n \over \spa1.n} \biggr)^{2 h - 4}
M_n^\tree(1^-, 2^-, 3^+, \ldots, n^+)\,,
\label{SWIforMHV}
\end{eqnarray}
where $a^h$ represents particle $a$ carrying helicity $h$, while on
the right hand side of this identity, we have a pure graviton tree
amplitude. For gravitons, $h$ takes on the values of $\pm2$. Similarly,
for gravitinos, it takes on the values $\pm 3/2$, and so forth.
(However, for simplicity of notation, in general, for graviton
amplitudes we keep only the $\pm$ label on each graviton leg.)  At
tree level this identity holds even in non-supersymmetric theories,
but at loop level they hold only in supersymmetric theories.  In the
unitarity cuts, this identity is rather useful, giving us a simple
means of separating the cancellations due to supersymmetry from those that
happen with no supersymmetry.  A similar identity holds for the
\MHVbar\ amplitudes, except that angle brackets are replaced with
square ones.

%%%%%%%%% FIGURE %%%%%%%%%%%%%%%
\begin{figure}[t] 
\centerline{\epsfxsize 2.0 truein \epsfbox{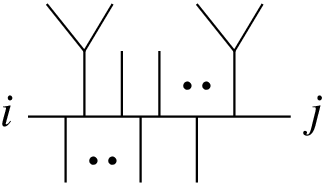}}
{\vskip -.3 cm }
\caption[a]{\small A gravity Feynman diagram.  Individual diagrams 
display bad behavior as $z \rightarrow \infty$.}
\label{GravityFeynmanFigure}
\end{figure}
%%%%%%%%%%%%%%%%%%%%%%%%%%%%%%%%

\subsection{Cancellations in tree amplitudes}

One clue pointing to improved high energy behavior of gravity
scattering amplitudes comes from considerations of on-shell recursion
relations~\cite{BCFRecursion,BCFW} for gravity 
amplitudes~\cite{GravityRecursion, CachazoLargez}. 
These recursion relations are
constructed by deforming the amplitude via a complex shift of
momenta. For example, we may shift two of the momenta, say those of
legs $i$ and $j$,
 \begin{equation}
k^\mu_i \rightarrow k^\mu_i(z) = k^\mu_i - {z\over2}\sand{i}.\gamma^\mu.j \, , 
  \hskip 1 cm 
k^\mu_j \rightarrow k^\mu_j(z) = k^\mu_j + {z\over2}\sand{i}.\gamma^\mu.j \, , 
\label{ShiftMomentum}
\end{equation} 
where $z$ is a complex parameter.  In terms of spinor variables, the
deformation (\ref{ShiftMomentum}) is equivalent to a shift of 
$\tlambda_i$ and $\lambda_j$,
\begin{equation}
\tlambda_i \rightarrow \tlambda_i - z\, \tlambda_j \,,  \hskip 2 cm 
\lambda_j \rightarrow \lambda_j + z\, \lambda_i \,.
\label{SpinorShift}
\end{equation} 
We will refer to this as an ``$\Shift{i}{j}$ shift''.  The above shift
maintains both momentum conservation and leaves legs $i$ and $j$ on
shell, deforming the amplitude $M_n^\tree(z)$ to become $z$ dependent.
As described in refs.~\cite{BCFW}, if a shifted tree amplitude vanishes
for large $z$, it can be written as a sum over
simple poles in $z$, giving rise to an on-shell recursion relation.

Taking $z$ in \eqn{ShiftMomentum} large, corresponds to taking the
momenta $k_i(z)$ and $k_j(z)$ large in a complex direction, which may
be interpreted as a particular high-energy limit. As discussed in
refs.~\cite{GravityRecursion, CachazoLargez}, for
the case of gravity, it is not at all obvious from Feynman diagrams
that the shifted amplitude, $M^\tree_n(z)$, will in fact vanish for
large $z$. Consider the covariant single gravity Feynman diagram displayed in
\fig{GravityFeynmanFigure}, assuming we have $m$ vertices and thus
$m-1$ propagators along the line connecting legs $i$ and $j$.  Each of
these vertices scale as $z^2$ since two shift momenta appear at each
vertex.  The shifted propagators scale as $1/z$, as can be easily
checked using the fact that $\sand{i}.\gamma^\mu.j$ is null. If the
shifted legs $i$ and $j$ are of negative and positive
helicity respectively, the polarization tensors for these legs give factors of
$1/z^2$ each, as can be seen by applying the shift (\ref{SpinorShift})
to \eqn{GravitonPol}.  Combining the various factors of $z$ gives us
an overall scaling for the diagram of ${z^{m-3}}$ which for $m\ge3$ is
badly behaved as $z \rightarrow \infty$.  In an $n$-point amplitude
Feynman diagrams may have up to $m=n-2$ vertices, so the worst behaved
diagrams scale as $z^{n-5}$ at large $z$.  This may be contrasted to
the behavior of Yang-Mills Feynman diagrams.  In this case, there is at
most one momentum at each vertex, leading to a large $z$ scaling in any
diagram no worse than $1/z$, after incorporating the behavior of the
polarization vectors (\ref{GluonPol}) of legs $i$ and $j$.  Thus,
gravity would appear to be much worse behaved at large $z$ than gauge
theory, in line with the standard statements that gravity is badly behaved
in the ultraviolet.  For $M_n^\tree(z)$ to vanish as $z\rightarrow
\infty$ there need to be hidden cancellations not apparent within
individual Feynman diagrams.

Indeed this is the case: various studies of on-shell recursion
relations support this~\cite{GravityRecursion} and, more recently,
Benincasa, Boucher-Veronneau and Cachazo~\cite{CachazoLargez} have
given a general proof for the vanishing of $n$-graviton amplitudes
under a $\Shift{-}{+}$ shift, where the ``$-$'' and ``$+$'' labels
refer to the helicities of the shifted legs in \eqn{ShiftMomentum}.
In particular, for MHV amplitudes, supersymmetry Ward identities give
the scaling of MHV graviton amplitudes for generic shifts which we
have collected in \tab{ShiftZTable}.  This pattern has been
conjectured to hold for any $n$-graviton tree
amplitude~\cite{NoTriangleSixPt}.  By following similar reasoning as
in ref.~\cite{CachazoLargez}, for the $\Shift{-}{+}$ shift, we have
confirmed that under a $\Shift{+}{-}$ shift $n$-graviton amplitudes
behave as $z^6$ for any helicity configuration. We have also confirmed
that \tab{ShiftZTable} is correct for {\it all} graviton helicity
configurations through at least ten points, by numerically evaluating
complex deformations of amplitudes constructed via on-shell recursion.
The cases of $\Shift{-}{-}$ and $\Shift{+}{+}$ shift remain to be
proven for $n>10$.

%%%%%%%%%%% TABLE %%%%%%%%%%%%%%%%%
\def\hs{\null \hskip .2 cm \null }
\begin{table*}[t]
\vspace{0.5cm}
\begin{tabular}{|l|c|c|c|c|}
\hline\hline
\hs helicity of shifted legs \hs & \hs $\Shift{-}{+}$\hs & \hs 
$\Shift{+}{-}$\hs & \hs $\Shift{+}{+}$\hs & \hs $\Shift{-}{-}$ \hs \\
\hline
\hs large $z$ scaling \hs	 & $z^{-2}$    	  & $z^{6}$         & $z^{-2}$        & $z^{-2}$ \\
\hline\hline
\end{tabular}
\vspace{0.2cm}
\caption{The leading scaling in $z$ of $n$-graviton amplitudes under
the shift in \eqn{ShiftMomentum}. The ``$+$'' and ``$-$'' labels refer
to the helicities of the shifted legs. 
\label{ShiftZTable}} 
\end{table*}
%%%%%%%%%%% TABLE %%%%%%%%%%%%%%%%%

For MHV amplitudes, the BGK form (\ref{BGKMHV}) (or
\eqn{ThreeFourPointTree} for three or four points) provides a rather
simple means to confirm the pattern in \tab{ShiftZTable}.  Under a
$\Shift{1}{n}$ shift, ignoring the overall $\spa{a}.{b}^8$, each term
in the sum in \eqn{BGKMHV} behaves as $1/z^2$ as $z \rightarrow \infty$
 because $\spa1.n$ is
unshifted, $\spa2.n \rightarrow \spa2.n + z \spa2.1$, and there is a
cancellation of $z^{n-3}$ between the products in the numerator
and denominator.  The pattern in \tab{ShiftZTable} then follows,
depending on whether the negative helicity legs $a$ or $b$, in
\eqn{BGKMHV}, correspond to legs $1$ and $n$.  Note that had we chosen
to shift other legs, the large $z$ behavior would not be manifest.  Indeed, 
each term would appear to have a worse behavior.  However, in the sum,
non-trivial cancellations between terms restore the scaling given in
\tab{ShiftZTable}.  

Interestingly, these cancellations provide some useful insight into
the ultraviolet properties of gravity.  The poor ultraviolet
properties of gravity are usually ascribed to its two-derivative
coupling, as well as the appearance of an infinite number of contact
terms.  The very existence of on-shell recursion relations calls into
question these standard arguments.  In particular, the better than
expected behavior, under large $z$ deformations, suggests an improved
high energy behavior, since this corresponds to a limit where momenta
are becoming large in certain complex directions.  Very remarkably,
on-shell recursion relations allow us to obtain all tree amplitudes in
gravity theories starting solely from the three-point vertices.  The
higher-point vertices that occur with conventional formulations are
unnecessary.  This is undoubtedly tied to the earlier realization
that the higher-point vertices of gravity follow from principles of
gauge and Lorentz covariance, without providing any additional
dynamical information to the scattering
amplitudes~\cite{FeynmanHigherPoint}.

%%%%%%%%%%%%%%%%%%%%%%%%%%%%%%
\section{One-loop gravity amplitudes and power counting}
\label{OneLoopSection}

Generalized unitarity allows us to use the tree-level
amplitudes, described in the previous section, directly to obtain properties of
loop-level amplitudes~\cite{GeneralizedUnitarity}.  We use
four-dimensional amplitudes in the cuts, allowing us to apply powerful
spinor methods~\cite{SpinorHelicity,TreeReview}.  Thanks to new
developments in evaluating integrals at one loop~\cite{BCFUnitarity,
BBCFOneLoop, OPP, ZoltanDdim, NoTriangleSixPt}, and in particular the
formalism of ref.~\cite{Forde}, we will be able to translate
straightforwardly from power counting in cuts to power counting in
Feynman integrals.

\subsection{Bubble-triangle cancellations}

When starting from Feynman diagrams, a one-loop $n$-point amplitude is
composed of a sum over loop integrals with up to $n$ propagators
carrying loop momentum.  Consider the generic case of an $m$-gon
integral with $m$ such propagators,
\begin{equation}
I_m = \int {d^D l \over (2 \pi)^D} \; {P_m(l) \over 
 l^2 (l - K_1)^2 (l - K_1 - K_2)^2 \cdots (l- \sum_{j=1}^{m-1} K_j)^2} \,,
\label{GeneralPolynomial}
\end{equation}
where the $K_i$ are sums over external momenta and $P_m(l)$ is a
numerator polynomial in the loop momentum $l$.  Because the integrals
can contain infrared and ultraviolet divergences, we use dimensional
regularization~\cite{HV}, analytically continuing the loop momentum integral
to $D= 4-2\e$ dimensions. (For the
supersymmetric case we use the four-dimensional helicity scheme
variant~\cite{FDH}, which is a relative of dimensional
reduction~\cite{DimRed}.)

%%%%%%%%%% FIGURE %%%%%%%%%%%%%%%
\begin{figure}[t]
\centerline{\epsfxsize 4.3 truein \epsfbox{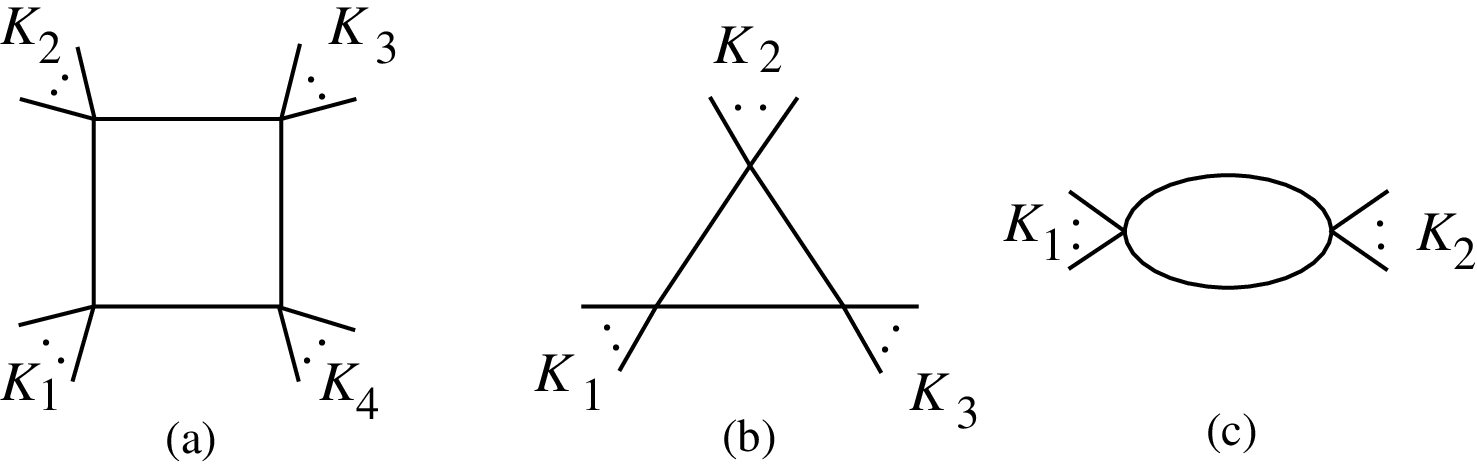}}
{\vskip -.3 cm }
\caption[a]{\small The different types of box, bubble and triangle
integrals are characterized by the momenta $K_i$ at each corner.  The
$K_i$ are sums of momenta of external particles.  Non-trivial
cancellations of numerator loop momentum occur within a given integral
type.}
\label{IntClassesFigure}
\end{figure}
%%%%%%%%%%%%%%%%%%%%%%%%%%%%%%%%%

Given an $m$-gon integral with $p$ powers of loop momentum in the
numerator and massless propagators there is a standard procedure ---
known as Brown-Feynman or Passarino-Veltman reduction --- which
reduces the integral to a basis set of scalar box, triangle and bubble
integrals, with no powers of loop momentum in their
numerators~\cite{PV,DimRegPV}.  The essential trick is to trade one
power of loop momentum in the numerator for a difference of inverse
propagators. For example, if $1/l^2$ and $1/(l-k_1)^2$ are two
propagators and $k_1^2=0$, we can rewrite,
\begin{equation}
2 l \cdot k_1 = l^2 - (l - k_1)^2 \,.
\end{equation}
By canceling these against propagators we obtain a difference of
integrals with one less propagator and one less power of loop
momentum in their numerators. If instead of $2 l \cdot k_1$ the
numerator factor was already an inverse propagator, say $l^2$,
we could cancel two powers of loop momentum.  Generically,
both types of terms will appear, with the former leading
to a worsening of the power counting in the reduced integrals.
Although this reduction procedure cannot alter the overall ultraviolet
behavior of the gravity amplitude, it does alter the power counting of
individual integrals.  (Cancellations
between integrals restore the proper overall behavior.)

In gauge theories, in Feynman gauge, one-loop $m$-gon integrals carry
up to $m$ powers of loop momentum in the numerator, due to the
one-derivative coupling, and $2m$ powers in the denominator from the
propagators.  Thus with increasing number of legs the ultraviolet
behavior of $m$-gons gets softer in the ultraviolet.  One step of the
integral reductions trades one power of loop momentum in the numerator
for two powers in the denominator giving $(m-1)$-gon integrals with no
worse than $(m-1)$ powers of loop momentum in the numerators and
$2(m-1)$ powers in the denominators from the propagators.  Carrying
out the chain of reduction down to the triangle level, we encounter
triangle integrals with no more than three powers of loop momentum
in their numerators. Similarly, carrying out the reduction down to the
bubble integrals, we encounter no more than two powers of loop momentum in the
numerator.  This matches the power counting for the Feynman diagrams
directly containing bubble and triangle loop integrals.

In contrast, gravity $m$-gon integrals may carry up to $2m$ powers of
loop momentum in their numerators, given the two-derivative couplings
of the theory, and $2m$ powers in the denominator, since propagators
are quadratic in the loop momentum.  Since the power of loop momentum
from vertices matches the power from the propagators, the leading
ultraviolet behavior of $m$-gon integrals is independent of $m$.
Starting from $P_m(l)$ with degree $2m$, in carrying out the chain of
integral reductions, we obtain $(m-r)$-point integrals with numerator
polynomials of degree $2m-r$ (where $r> m-2$).  In particular, in an
$n$-point gravity amplitude, under integral reductions, starting from
an $n$-gon integral in the amplitude, we would obtain triangle
integrals with numerator polynomials $P_3(l)$ of degree $n+3$ and
bubble integrals with numerator polynomials $P_2(l)$ of degree
$n+2$. That is, the triangle and bubble integrals encountered in the
chain of integral reductions would appear to have a worse power
count than their parent $m$-gon integrals.

The central observation of the present article is that non-trivial
cancellations lead to a very different pattern for the power counting
of individual integrals. We shall find that the maximum degree is six
in the triangle case and four in the bubble case, {\it i.e.}
\begin{equation}
P_2(l)\sim l^4\,,\hskip 2 cm P_3(l)\sim l^6\, , 
\label{trianglebubblecancellation}
\end{equation}
independent of the number of external legs.

To investigate the power counting at the triangle level, our approach
will be equivalent to reducing all $(m>3)$-point integrals to scalar
box integrals and tensor triangle integrals, with no further integral
reductions, as any further reductions would lower the degrees of the
triangle numerator polynomials $P_3(l)$.  Similarly, to perform the
power counting at the bubble level, we will effectively reduce all
$m>2$ integrals to scalar boxes, scalar triangles, and tensor bubble
integrals, again with no further integral reductions that would lower
the degree of the bubble numerator polynomials $P_2(l)$.  This will
allow us to demonstrate the existence of non-trivial cancellations in
one-loop gravity amplitudes.
We focus in our analysis on the tensor bubble and triangle 
integrals, since this allows us to identify cancellations straightforwardly.

To obtain a complete evaluation of the amplitude one continues the
reduction procedure until only known scalar integrals remain.  In this
way, any dimensionally regularized one-loop amplitude can be expressed
as a linear combination of basis scalar integrals multiplied by
rational coefficients~\cite{DimRegPV},
\begin{equation}
A_n^\oneloop = \sum_j a_j I_1^j +
\sum_j b_j I_2^j + \sum_j c_j I_3^j  + \sum_j d_j I_4^j
+ \hbox{finite rational}\,,
\label{BasisIntegrals}
 \end{equation}
where $I_2^j, I_3^j$ and $I_4^j$ are scalar bubble, triangle and box
integrals, respectively and $b_j, c_j, d_j$ their rational
coefficients. This structure near four dimensions, where no integrals
beyond  box integrals appear in the basis, is a 
general property.  In massive theories we also obtain ``cactus'' or
one-point contributions, $I_1^j$, but with dimensional regularization
these are set to zero for massless particles, as in the cases
discussed in this paper.  An unwanted side effect of carrying out the
complete reduction to the basis of scalar integrals, is that the power
counting becomes more obscure, due to the way that dimensional
regularization sets power divergences to zero. For this reason,
to carry out power counting, it is much more convenient  to identify
cancellations prior to eliminating tensor triangle and bubble
integrals.\footnote{The information on which
tensor triangle or bubble integrals appear in the amplitude remains
encoded in the coefficients of the basis scalar integrals via powers
of the spurious poles as well as particular numerical factors.}  Besides the
scalar integrals, there are also finite rational terms in
\eqn{BasisIntegrals} which arise when the $-2\eps$ dimensional
components of loop momentum interfere with a $1/\eps$ ultraviolet
singularity.  We defer the study of the finite rational terms to the
future.

\subsection{Relation to one-loop finiteness}

It is useful to compare the above cancellations to the well known
ultraviolet cancellations underlying the one-loop
finiteness~\cite{DeWitt,tHooftVeltmanGravity} of pure Einstein
gravity.  In the language of Lagrangian counterterms one-loop pure
gravity is finite because all potential counterterms vanish on shell
or equivalently by the equations of motion.  The only three potential
one-loop counterterms are: $R^2$, $R_{\mu\nu}^2$, and
$R_{\mu\nu\sigma\rho}^2$, where $R$ and $R_{\mu\nu}$ are the Ricci
scalar and tensor, while $R_{\mu\nu\sigma\rho}$ is the Riemann
tensor. The first two of these potential counterterms can be removed
by field redefinitions and vanish on shell, while the Gauss-Bonnet
theorem, in four dimensions, allows the squared Riemann tensor to be
expressed as a linear combination of the other two, so it too is not a
viable counterterm. This delays the divergences until two loops.

In the language of $S$-matrix elements, the ultraviolet finiteness is
due to a cancellation between the coefficients of distinct bubble
integrals, depending on different kinematic invariants.  If the
amplitude is fully reduced to the basis of scalar
integrals~(\ref{BasisIntegrals}), the ultraviolet divergences are
found in bubble integrals, while the triangle and box integrals are
infrared divergent, but ultraviolet finite. The explicit form of the
integrals may be found, for example, in Appendix~I of
ref.~\cite{Fusing}.  In particular, the scalar bubble integral is,
\begin{equation}
 I_2(s) = {i \over (4 \pi)^{2-\eps} }
 {\Gamma(1+\eps) \Gamma^2(1-\eps) \over \Gamma(1-2\eps) }
   \biggl( {1\over \eps} - \ln(-s) +2 \biggr) + \Ord(\eps)\,,
\label{BubbleIntegral}
\end{equation}
where $1/\eps$ is an ultraviolet divergence. 
To see how the cancellation arises 
we may use the four-point amplitudes in pure
gravity, as explicitly computed by Dunbar and
Norridge~\cite{DunbarNorridge}.  For the pure gravity one-loop
amplitude $M_4^\oneloop(1^-, 2^-, 3^+, 4^+)$ the 
bubble integrals enter in the combination,
\begin{equation}
I_2(s_{14}) - I_2(s_{13}) \sim \ln(s_{14}/s_{13}) \,,
\end{equation}
and the ultraviolet divergences cancel between the two integrals.  If
matter is added, the theory will no longer be one-loop
finite~\cite{tHooftVeltmanGravity, DeserMatter}.  Indeed, the
four-scalar amplitude, for example, diverges since the bubble integral
divergences no longer cancel~\cite{DunbarNorridge}.

In contrast to the above cancellations between integrals, our study of
cancellations concerns triangle and bubble integral functions within
the given class specified by the external momenta at each corner, as
displayed in \fig{IntClassesFigure}. We shall find that the power
counting within the individual classes is better than naively expected
and anticipate that this will have important consequences at higher
loops, in much the same way as the no-triangle hypothesis constrains
the higher-loop ultraviolet behavior in $\NeqEight$ supergravity
via unitarity~\cite{Finite}.

\subsection{Large momentum scaling in integrals}

%%%%%%%%%% FIGURE %%%%%%%%%%%%%%%
\begin{figure}[t]
\centerline{\epsfxsize 4.7 truein \epsfbox{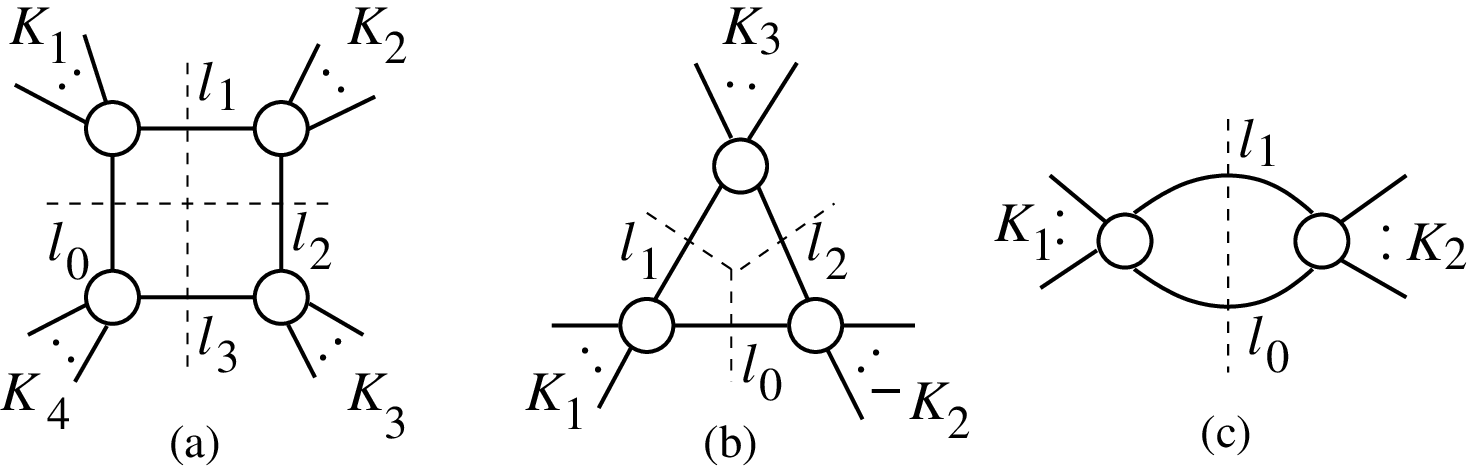}}
{\vskip -.3 cm }
\caption[a]{\small The (a) quadruple, (b) triple and (c) ordinary double
cut. (The minus sign in the definition of $K_2$ in the triple cut follows
the conventions of ref.~\cite{Forde}.)  We take $l_0 \equiv l$.
}\label{CutsFigure}
\end{figure}
%%%%%%%%%%%%%%%%%%%%%%%%%%%%%%%%%

First consider box integrals.  The coefficients of these integrals are
most easily determined from quadruple cuts~\cite{BCFUnitarity}, as
shown in \fig{CutsFigure}(a). If we replace the four Feynman
propagators by on-shell delta functions we obtain an integral of the
form,\footnote{Since these integrals are in general infrared divergent
one need to replace $d^4l$ with $d^D l$, but this turns out to have no
effect on determining the integral coefficients for $D \rightarrow
4$; this difference, however, does affect the finite rational remainder
in \eqn{BasisIntegrals}~\cite{Fusing}.}
\begin{equation}
\int {d^4 l} \; \delta(l^2) \,\delta((l-K_1)^2)\, \delta((l-K_1-K_2)^2) \,
          \delta((l+K_4)^2) \, M_{(1)}^\tree \, M_{(2)}^\tree \, M_{(3)}^\tree
                \, M_{(4)}^\tree \,,
\end{equation}
where the $M_{(i)}^\tree$ correspond to the four tree amplitudes
sitting at the corners of the box in \fig{CutsFigure}(a).  The
quadruple cut conditions freeze the loop integral in four dimensions,
since there are four on-shell conditions,
\begin{equation}
l^2 = 0 \,, \hskip 1 cm 
(l-K_1)^2 = 0 \,, \hskip 1 cm 
(l-K_1-K_2)^2 = 0 \,, \hskip 1 cm 
(l+K_4)^2 = 0 \,,
\end{equation}
allowing us to solve for the loop momentum directly in terms of the
external momenta. (Further details and
examples of calculations with quadruple cuts may be found in
refs.~\cite{BCFUnitarity,OnShellReview}.)
The coefficient in front of the integral is then
simply given by substituting the solution for $l$ into the product of
tree amplitudes.  Although it is an efficient means of obtaining the
rational coefficients of all basis box integrals in
\eqn{BasisIntegrals}, this has an unwanted side effect: once the
quadruple cut conditions are imposed and the solved loop momentum
inserted in, we can no longer perform power counting as the loop momentum has
disappeared from the numerator polynomials.  Similar considerations
also prevent us from using generalized cuts to power
count straightforwardly the $m$-gon integrals with $m>4$.

We therefore turn to triangle and bubble functions.  These functions may
be determined from triple and double cuts, as shown in \fig{CutsFigure}(a)
and (b).  With these cuts, there remain unfixed degrees of freedom in 
the loop integrals, allowing us to count powers of the loop momentum, 
using the parameters describing these.

Consider then the triangle integrals and the triple cuts which
determine them.  In this case, there are three cut conditions but four
components of loop momentum, leaving one unconstrained degree of
freedom.  Performing a triple cut as shown in
figure~\ref{CutsFigure}(b), allows us to isolate the triangle integral
specified by the external legs at each corner.  The triple cut is of
the form,
\begin{equation}
\int {d^4 l} \; \delta(l^2) \, \delta((l-K_1)^2) \, \delta((l-K_2)^2) \, 
          M_{(1)}^\tree \, M_{(2)}^\tree \, M_{(3)}^\tree \,. 
\label{TripleCut}
\end{equation}
Following the construction of
ref.~\cite{Forde}, we parameterize the cut loop momenta in terms of
the single unconstrained parameter $t$, 
\begin{eqnarray}
l^{\mu}_i=\alpha_{i2} K_1^{\flat,\mu} +
\alpha_{i1} K_2^{\flat,\mu} +
\frac{t}{2}\Bmp{K^{\flat,-}_1|\gamma^{\mu}|K^{\flat,-}_2}
+\frac{\alpha_{i1}\alpha_{i2}}{2t}
\Bmp{K^{\flat,-}_2|\gamma^{\mu}|K^{\flat,-}_1}\,,
\label{TripleCutMomentum}
\end{eqnarray} 
where $i = 0,1,2$ corresponds to the three cut lines of the triangle
as shown in \fig{CutsFigure}(b), with $l_0 \equiv l$.
The basis momenta are, 
\begin{eqnarray}
K_1^{\flat,\mu}=\frac{K^{\mu}_1-(S_1/\gamma)K_2^{\mu}}{1-(S_1S_2/\gamma^2)}\,,
\hskip 2 cm  K_2^{\flat,\mu}=
\frac{K_2^{\mu}-(S_2/\gamma)K^{\mu}_1}{1-(S_1S_2/\gamma^2)} \,,
\label{eq:def_gamma_pm}
\end{eqnarray} 
$S_1 = K_1^2$, $S_2 = K_2^2$ and $\gamma$ has two solutions given by, 
\begin{eqnarray}
\gamma_{\pm}=(K_1\cdot K_2)\pm\sqrt{\Delta}\,,
\hskip 1.5 cm \Delta=(K_1\cdot K_2)^2-K_1^2K_2^2\,.
\label{eq:def_delta}
\end{eqnarray} 
The $\alpha_{ij}$ parameters are functions of the $K_i$ whose explicit
form may be found in Appendix~A of ref.~\cite{Forde}.  This
parameterization is equivalent to the one used earlier in
ref.~\cite{OPP}.  

With this momentum parametrization, the spinors depending on the 
loop momenta are,
\begin{eqnarray}
\langle l_i^-|
&=&t\,\langle K_{1}^{\flat,-}|+\alpha_{i1}\,\langle K_{2}^{\flat,-}|,
\nonumber\\
\langle l_i^+|
&=&\frac{\alpha_{i2}}{t}\,\langle K_{1}^{\flat,+}|+\langle K_{2}^{\flat,+}|,
\end{eqnarray}
so that inner products involving
these are given by, 
\begin{eqnarray}
\spa{l_i}.{a} &=& 
t\, \spash{K_{1}^{\flat}}.{a} + \alpha_{i1}\spash{K_{2}^{\flat}}.a\,,
\hskip 2.00 cm 
\spb{l_i}.{a} =
 \frac{\alpha_{i2}}{t}\, \spbsh{K_{1}^{\flat}}.a +\spbsh{K_{2}^{\flat}}.{a}\,, 
\nonumber \\
\Bmp{l_0\,l_1} &=& -t\,\frac{S_1}{\gamma}\Bmp{K^{\flat}_1\, K^{\flat}_2}\,,
\hskip 3.1 cm 
\left[l_0\,l_1\right]=-\frac{1}{t}\,[K^{\flat}_2\,  K^{\flat}_1]\,, 
\nonumber\\
\Bmp{l_0\,l_2} &=& -t\,\Bmp{K^{\flat}_1\, K^{\flat}_2} \,,
\hskip 3.5 cm 
\left[l_0\,l_2\right]=-\frac{1}{t}\,\frac{S_2}{\gamma}
                  [K^{\flat}_2\, K^{\flat}_1] \, ,
\nonumber\\
\Bmp{l_1\,l_2}&=& -t\,\left(1-\frac{S_1}{\gamma}\right)
              \Bmp{K^{\flat}_1\,K^{\flat}_2} \,,
\hskip 1.6 cm 
\left[l_1\,l_2\right]= \frac{1}{t}\,\left(1-\frac{S_2}{\gamma}\right)
                   [K^{\flat}_2\,K^{\flat}_1] \,,
\label{SpinorsTripleCut}
\end{eqnarray}  
where each has a simple scaling in $t$.  There is an arbitrariness
in the scaling in the spinors since one can move powers of $t$ between
the angle and square brackets, leaving \eqn{TripleCutMomentum} the
same. Here we use the same choice as ref.~\cite{Forde}.  For unitarity
cuts this arbitrariness cancels because the opposite helicity spinors
carrying the cut loop momenta can always be paired up.

In ref.~\cite{Forde} these formul\ae\ were used to develop an
efficient method for extracting the coefficients of the scalar
triangle functions in the basis of integrals~(\ref{BasisIntegrals}).
Instead of evaluating integral coefficients, here we use this approach
to perform power counting in each class of triangle and bubble
integral functions, automatically accounting for all contributions of
integral reductions of higher-point integrals.  Using the cut
expressions it is then a simple matter to determine the maximum
numbers of loop momenta which occur in the numerators. This is done
simply by making the replacements of
\eqns{TripleCutMomentum}{SpinorsTripleCut} into the product of three
tree amplitudes in \eqn{TripleCut} and then determining the maximum
power of $t$ in the limit that $t$ becomes large.  From
\eqn{TripleCutMomentum}, each power of $t$ corresponds to a power of
$l$ in the numerator.  This gives us the maximum tensor triangle
integral that can occur had we carried out a Passarino-Veltman
integral reduction on the one-loop Feynman diagrams down to triangles.

Any terms that scale as $t^{-m}$ with $m>0$ as $t \rightarrow \infty$
in the cut will not contribute to the triangle integrals. For example,
a scalar box integral present in the triple cut will scale as $1/t$
and will drop out in the large $t$ limit.  If the entire triple cut scales as
an inverse power of $t$, then there is no triangle contribution at
all.

Next consider bubble integrals.  The behavior of the bubble integrals
can be obtained from the two-particle cuts, as shown in
figure~\ref{CutsFigure}(c).  Following ref.~\cite{Forde}, the
two-particle cuts leave two unconstrained parameters, which we label
$y$ and $t$, allowing the cut loop momentum to be parameterized as,
\begin{eqnarray}
l_0^{\mu}&=&y K_1^{\flat,\mu}+\frac{S_1}{\gamma}(1-y)\chi^{\mu}
+\frac{t}{2}
\Bmp{\Kfm1|\gamma^{\mu}|\chi^-}
+\frac{S_1}{2\gamma}\frac{y}{t}(1-y)\Bmp{\chi^-|\gamma^{\mu}|\Kfm1} \,,
\label{BubbleCut4Vector}
\end{eqnarray}
where $l_0 \equiv l$.
Here $\chi$ is an arbitrary null momentum and
\begin{eqnarray}
K_1^{\flat,\mu}=K_1^{\mu}-\frac{S_1}{\gamma}\chi^{\mu},
\end{eqnarray}
where $\gamma_\pm =\Bmp{\chi^\pm|\s K_1|\chi^{\pm}}$.

The spinors depending on the loop momentum are,
\begin{eqnarray}
\langle l_0^-|&=&t\,\langle
K_{1}^{\flat,-}|+\left(1-y\right)\,\frac{S_1}{\gamma}\langle
\chi^{-}|\,, \hskip 1.7 cm 
\langle l_0^+|=\frac{y}{t}\,\langle K_{1}^{\flat,+}|+\langle \chi^{+}| \,, \nn\\
\langle l_1^-|
&=&\langle K_{1}^{\flat,-}|-\frac{y}{t}\,\frac{S_1}{\gamma}\langle \chi^{-}|\,,
\hskip 2.9 cm 
\langle l_1^+|
=\left(y-1\right)\,\langle K_{1}^{\flat,+}|+t\,\langle \chi^{+}|\,, \hskip .5 cm
\end{eqnarray}
so that the inner products involving these are,
\begin{eqnarray}
\spa{l_0}.a&=&t\,\spa{K_{1}^{\flat}}.a+\left(1-y\right)\,\frac{S_1}{\gamma}\spa{\chi}.a\,,\hskip0.6cm
\spb{l_0}.a\,=\frac{y}{t}\,\spb{K_{1}^{\flat}}.a+\spb{\chi}.a\,,\nonumber\\
\spa{l_1}.a&=&\spa{K_{1}^{\flat}}.{a}-\frac{y}{t}\,\frac{S_1}{\gamma}\spa{\chi}.a\,,
\hskip2.0cm
\spb{l_1}.{a}=\left(y-1\right)\,\spb{K_{1}^{\flat}}.a+t\,\spb{\chi}.a\,,\nonumber\\
\spa{l_0}.{l_1}&=&{S_1\over \gamma}\Bmp{\Kf1\chi}\,,\hskip4.0cm \spb{l_0}.{l_1}=[\Kf1\chi]\,.
\label{DoubleCutMomDef}
\end{eqnarray}
In ref.~\cite{Forde} these were used to develop a method to extract
the value of any bubble integral coefficient.  Here we use it to
determine the maximum number of powers of loop momenta that can occur
in any bubble integral, simply by taking $y$ large and determining the
maximum power of $y^2$.  From \eqn{BubbleCut4Vector}, we see that in
the large $y$ limit each power of $y^2/t$ corresponds to an additional
power $l$ that can occur in the numerator of the bubble integrals.  In
the formalism of ref.~\cite{Forde} the coefficients of bubble
integrals also receive contributions from triple-cut terms.  However,
the contributions of such terms do not alter our conclusions on the
general scaling behavior of the bubble terms. Therefore, as we can
always infer the overall scaling behavior of the bubble terms purely
from the contributions of the two-particle cut terms, we do not need
to discuss these terms further.  If a two-particle cut scales as
$(y^2/t)^{-m}$, with $m>0$, then there is no bubble contribution at
all.  In a similar manner to that of the triple-cut case there is an
arbitrariness in the overall scaling of these spinor inner products
which cancels in the cuts.

%%%%%%%%%%%%%%%%%%%%%%%%%%

\section{Power counting of triangle and bubble integrals}
\label{PowerCountingSection}

In this section, we apply the formalism described in the previous
section to perform power counting on bubble and triangle integrals in
one-loop pure gravity amplitudes.  First we work through some examples
at six points, comparing to the earlier results obtained in
$\NeqEight$ supergravity~\cite{OneloopMHVGravity,NoTriangleSixPt}.
Similarly we consider the other ${\cal N}$-extended supergravities and
observe that theories with ${\cal N}\ge 5$ should be
``cut-constructible'', if the observed cancellations are universal. We
then present results for an arbitrary number of legs but for limited
classes of contributions.  We also numerically analyze the power
counting for {\it all} helicity configurations up to ten points in
pure gravity.  Some further results can be found in
\sect{CancellationSection}, where we outline a proof of the scaling
for the case when the two cut lines of each tree amplitude appearing
in the cuts are of opposite helicity.

\subsection{A six-point bubble example}

%%%%%%%%% FIGURE %%%%%%%%%%%%%%%
 \begin{figure}[t]
\centerline{\epsfxsize 3.6 truein \epsfbox{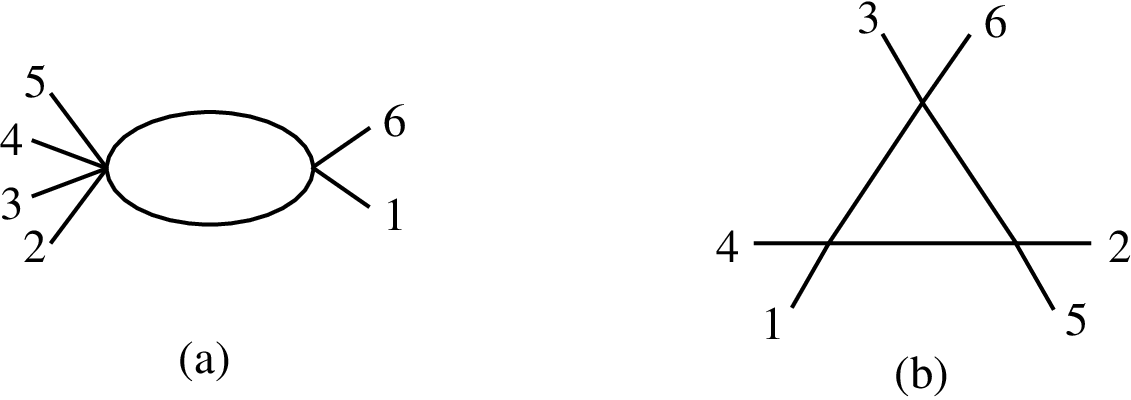}}
\caption[a]{\small Example bubble and triangle integrals appearing in the
six-point amplitude.}
\label{SixBubbleTriangleFigure}
\end{figure}
%%%%%%%%%%%%%%%%%%%%%%%%%%%%%%%%

\begin{figure}[b]
\caption[a]{\small An example of a Feynman diagram giving the
worst behaved contribution to the bubble integral in
\fig{SixBubbleTriangleFigure}(a).  Under a Passarino-Veltman reduction
a hexagon integral in gravity gives bubble integrals with up to eight
powers of loop momentum in the numerator. The dashed line represents
the channel used to evaluate the contribution of this diagram to the
bubble integral in \fig{SixBubbleTriangleFigure}(a) via the unitarity
cuts.  }
\centerline{\epsfxsize 1.4 truein \epsfbox{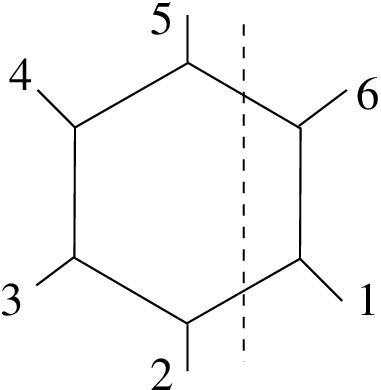}}
\label{HexagonCutFigure}
\end{figure}

As an instructive example, consider the six-point one-loop amplitude
$M_6^\oneloop(1^-,2^-, 3^+, 4^+,5^+,6^+)$ and the power counting of
the bubble integral shown in \fig{SixBubbleTriangleFigure}(a). 
Using the Passarino-Veltman reduction, the worst behaved contribution
to the bubble integral comes from the hexagon Feynman diagram displayed
in \fig{HexagonCutFigure}.  Because of the two-derivative coupling,
the hexagon Feynman diagram has 12 powers of loop momentum in 
the numerator.  Since each step of the reduction 
eliminates one power of loop momentum in the numerator, as well 
as one propagator, after four steps we obtain bubble integrals with
8 powers loop momentum in their numerators, prior to accounting
for any cancellations with other hexagon diagrams.

%%%%%%%%% FIGURE %%%%%%%%%%%%%%%
%%%%%%%%%%%%%%%%%%%%%%%%%%%%%%%%

To obtain the maximum power of loop momentum in
the bubble integral, including feed down from higher-point integrals,
we determine the scaling of the two-particle cut in
\fig{CutsFigure}(c) with $K_1 = k_2 + k_3 + k_4 +k_5$ and $K_2 = k_1
+ k_6$, selecting the desired integral. In this case, only two
helicity configurations of the cut legs contribute --- for the other
configuration one of
the tree amplitudes in the cut vanishes. Thus, the product of tree
amplitudes in the cut is,
\begin{eqnarray}
C^{\,\vspace{-0.1cm}{\rm gravity}\vspace{0.1cm}}_{61} 
&=& M_4^\tree(l_0^-, 6^+, 1^-, -l_1^+) \times
       M_6( -l_0^+, 2^-, 3^+, 4^+, 5^+, l_1^-) \nn \\
&& \null 
     + M_4^\tree( l_0^+, 6^+, 1^-, -l_1^-) \times
       M_6( -l_0^-, 2^-, 3^+, 4^+, 5^+, l_1^+) \,. \hskip .5 cm 
\label{GravityCut61}
\end{eqnarray}

First consider the behavior of the individual Feynman diagrams that
compose the tree amplitudes in the cut.  With the cut conditions used
to determine the integrals, the large $y$ scaling is similar to that
of Feynman diagrams under large $z$ scaling, described in
\sect{TreeSection}.  From \eqn{BubbleCut4Vector}, each power of loop
momentum in the numerator generically counts as one power of $y^2$.
Since $l^2 = 0$, each propagator, $1/(l - K_i)^2 \sim 1/(2\,l\cdot
K_i)$, counts as a power of $1/y^2$. The worst behaved contribution
then comes from diagrams with the maximal number of propagators and
vertices on each side of the cut, an example of which is the hexagon
shown in \fig{HexagonCutFigure}.  Diagrams with higher-point vertices
or with trees attached to the loop are subdominant in the scaling.
Since at six points there are a maximum of six vertices and four
propagators (not counting the cut ones) this gives us a large $y$
scaling of $(y^2)^{12}/ (y^2)^4 = (y^2)^8 $.  (The product of
polarization tensors of the cut legs have a canceling scaling in $y$.)
This corresponds to a bubble integral with up to eight powers of the
loop momentum $l^\mu$ in the numerator, matching the above counting
from Passarino-Veltman integral reductions.  Of course, neither of
these power counts account for any cancellation between the diagrams.
In order to see any such cancellations we need to use the explicit
form of the tree amplitudes in the cuts.

As a warm-up prior to evaluating the gravity cut (\ref{GravityCut61}),
it is useful to consider the cut analysis in Yang-Mills theory. The
corresponding unitarity cut with the external legs color ordered is
given by,
 \begin{eqnarray}
C_{61}^{\rm YM} & = &
A_4^{\tree}(l_0^-, 6^+, 1^-, -l_1^+) \times  
              A_6^{\tree}(-l_0^+,2^-,3^+,4^+,5^+,l_1^-) \nn \\
&& \hskip .5cm \null
+ A_4^{\tree}(l_0^+, 6^+, 1^-, -l_1^-) \times  
              A_6^{\tree}(-l_0^-,2^-,3^+,4^+,5^+,l_1^+)\,. 
\label{ymcut}
\end{eqnarray}
The first of these terms is given by,
\begin{equation} 
{i\,\spa{1}.{l_0}^4 \over 
     \spa{l_1}.{l_0} \spa{l_0}.{6} \spa{6}.{1} \spa{1}.{l_1}} \times
{i\,\spa{l_1}.2^4 \over  
     \spa{l_1}.{l_0}\spa{l_0}.2\spa2.3 \spa3.4 \spa4.5 \spa{5}.{l_1}}
 \sim y^2\,t \times y^2/t^3 \sim (l^\mu)^2\,,
\end{equation}
where we used the explicit forms of the MHV gluon
amplitudes~\cite{ParkeTaylor,TreeReview}.  To obtain the large-$y$ scaling
we use \eqn{DoubleCutMomDef}. Thus, we have a maximum of two
powers of loop momentum in the bubble integrals from the first term. 
 The second term in \eqn{ymcut} is similar,
\begin{equation} 
{i\,\spa{1}.{l_1}^4 \over 
     \spa{l_1}.{l_0} \spa{l_0}.{6} \spa{6}.{1} \spa{1}.{l_1}} \times
{i\,\spa{l_0}.2^4 \over  
     \spa{l_1}.{l_0}\spa{l_0}.2\spa2.3 \spa3.4 \spa4.5 \spa{5}.{l_1}}
\sim  {y^2 \over t^3} \times y^2\,t \sim (l^\mu)^2\,.
\end{equation}
Thus, we reproduce the well know result that bubble integrals
in gauge theory can have up to two powers of loop momentum in their
numerators, including the contribution from the reduction of 
higher-point tensor integrals.

The KLT relations (\ref{KLTFourPoint}) and (\ref{KLTSixPoint}) give us
a simple way to compare the gravity bubble integral power
counting directly to one for gauge theory and to identify cancellations. Using
the labels appearing in the gravity cut (\ref{GravityCut61}), we have from the
KLT relations,
\begin{eqnarray} 
M_4^{\tree}(l_0, 6, 1, -l_1) & = & -i s_{61}\, A_4^{\tree}(l_0, 6, 1, -l_1)
                                          A_4^{\tree}(l_0, 1, 6, -l_1)\,, \\
M_6^{\rm tree}(-l_0,2,3,4,5,l_1) &=&
 - is_{l_0 2}s_{45}\ A_6^{\tree}(-l_0,2,3,4,5,l_1)
\Bigl(s_{l_0 3}A_6^{\tree}(2,3,-l_0,5,4,l_1) \nn\\
&& \hskip 0.5 cm \null
\;+\;(s_{l_0 3}-s_{23})\ A_6^{\tree}(3,2,-l_0,5,4,l_1) \Bigr) 
\;+\; \Perm(2,3,4)\,, \hskip .7 cm 
\label{KLTSixCut}
\end{eqnarray} 
where the six-point expression is obtained from \eqn{KLTSixPoint},
by first relabeling  $(1,2,3,4,5,6) \rightarrow  (5,4,3,2, -l_0, l_1)$, then
using the invariance of the color-ordered gauge-theory tree amplitudes
under cyclic permutations and reversal of leg labels.

Using these expressions it is straightforward to evaluate the large $y$ scaling
in the gravity cut in \eqn{GravityCut61}. For the four-point
amplitudes appearing in the cut we have,
\begin{eqnarray} 
M_4^\tree(l_0^-, 6^+, 1^-, -l_1^+)  &=& -i s_{61} 
{i\spa{1}.{l_0}^4 \over\spa{l_1}.{l_0} \spa{l_0}.6 \spa 6.1 \spa1.{l_1}} \,
{i\spa{1}.{l_0}^4 \over \spa{l_1}.{l_0} \spa{l_0}.1 \spa 1.6 \spa6.{l_1}} \sim
y^4\,t^2\,, \nn \\
M_4^\tree(l_0^+, 6^+, 1^-, -l_1^-)  &=& -i s_{61} 
{i\spa{1}.{l_1}^4 \over\spa{l_1}.{l_0} \spa{l_0}.6 \spa 6.1 \spa1.{l_1}} \,
{i\spa{1}.{l_1}^4 \over \spa{l_1}.{l_0} \spa{l_0}.1 \spa 1.6 \spa6.{l_1}} \sim
y^4\, t^{-6}\,. \hskip 1. cm 
\label{FourPointScaling}
\end{eqnarray}
On the other side of the cut, for the six-point tree amplitudes, 
some terms appear to be worse 
behaved because of the explicit kinematic invariants carrying loop momentum 
appearing in the KLT expression~(\ref{KLTSixCut}).  However, there is a
non-trivial cancellation in such terms.
In particular, consider the following terms in the six-point amplitude in
\eqn{KLTSixCut} with helicity configuration
$M_6^{\tree}(-l_0^+,2^-,3^+,4^+,5^+,l_1^-)$:
\begin{eqnarray} 
&& \hspace{-0.5cm}s_{l_0 2}s_{l_0 3} 
A_6^{\tree}(-l_0^+,2^-,3^+,4^+,5^+,l_1^-) \nn \\
&& \hskip .7 cm \null \times 
   \Bigl(A_6^{\tree}(2^-,3^+,-l_0^+,5^+,4^+,l_1^-) 
     +  A_6^{\tree}(3^+,2^-,-l_0^+,5^+,4^+,l_1^-) \Bigr)  \nn \\
&& 
= s_{l_0 2}s_{l_0 3} {i\spa{l_1}.2^4 \over 
      \spa{l_1}.{l_0}\spa{l_0}.2 \spa2.3 \spa3.4 \spa 4.5 \spa{5}.{l_1}}
{i\spa{l_1}.2^4 \over \spa{l_0}.5\spa5.4 \spa 4.{l_1}\spa2.3 }
    \biggl( {1\over \spa{l_1}.2 \spa{3}.{l_0} } -
            {1\over \spa{l_1}.3 \spa{2}.{l_0} } \biggr) \nn \\
&& 
=  s_{l_0 2}s_{l_0 3}  {i\spa{l_1}.2^4 \over 
      \spa{l_1}.{l_0} \spa{l_0}.2 \spa2.3 \spa3.4 \spa 4.5 \spa{5}.{l_1}}
{i\spa{l_1}.2^4 \over \spa{l_0}.5\spa5.4 \spa 4.{l_1} }
     {\spa{l_1}.{l_0}
  \over \spa{l_1}.2 \spa{3}.{l_0} \spa{l_1}.3 \spa{2}.{l_0} } \nn \\
&& 
\sim y^4\,t^{-6}\,,
\label{SixPointScaling}
\end{eqnarray}
where we made use of the Schouten identity (\ref{Schouten}).  Note the
explicit factor of $s_{l_0 2}s_{l_0 3}$ in front of these terms, which
makes the expression appear badly behaved. However, there are
compensating reductions in the scaling. One reduction comes from $l_1$
and $l_0$ being non-adjacent in the color ordering of one of the
Yang-Mills factors.  In addition, there is a non-trivial cancellation
between the terms again reducing the scaling by another power of
$y^2$, giving the overall scaling of $y^4$.  It is not difficult to
check that all the remaining terms in \eqn{KLTSixCut} scale the same
way.  The net effect is that the six-point tree amplitudes appearing
in the cut (\ref{GravityCut61}) scale as,
\begin{eqnarray}
M_6^\tree( -l_0^+, 2^-, 3^+, 4^+, 5^+, l_1^-) &\sim&  y^4\,t^{-6}\,,\\
M_6^\tree( -l_0^-, 2^-, 3^+, 4^+, 5^+, l_1^+) &\sim&  y^4\,t^{2} \,.
\label{SixTreeScaling}
\end{eqnarray}
Combining this scaling with that of \eqn{FourPointScaling} gives us
that the two particle cut (\ref{GravityCut61}) scales as, 
\begin{equation}
C^{\,\vspace{-0.1cm}{\rm gravity}\vspace{0.1cm}}_{61} \sim y^8t^{-4} 
   \sim (l^\mu)^4 \,.
\end{equation}

We thus conclude that the maximum number of powers of loop momentum
that can appear in the numerator of the gravity bubble integral shown in
\fig{SixBubbleTriangleFigure}(a) is $(l^\mu)^4$.  This is significantly
fewer powers than the $(l^\mu)^8$ that is found from the
integral reduction of an individual hexagon gravity Feynman diagram
down to bubble integrals.  It is important to note the non-trivial
cancellation in \eqn{SixPointScaling} required to obtain this result:
only after the terms are combined do we find the better scaling behavior.

At this point we may compare to the known $\NeqEight$ supergravity
six-graviton MHV amplitude~\cite{OneloopMHVGravity}, in order to
separate the cancellations due to supersymmetry from those inherent to
generic gravity theories.  The supergravity case is quite similar to
the pure gravity case, except that we need to sum over the
contributions of the super-multiplet in the loop.  Thanks to the MHV
supersymmetry Ward identities (\ref{SWIforMHV}) this task is
straightforward.  In ${\cal N}=8$ supergravity, we must sum over the
contribution of the 256 different states of the theory.  These
correspond to one graviton, 8 gravitinos, 28 vectors, 56 spin 1/2
fermions, and 70 real scalars.  Using the supersymmetry Ward identity
(\ref{SWIforMHV}) and taking the contribution of the super-multiplet
into account by an overall factor $\rho_8$, we have,
\begin{eqnarray}
C^{\vspace{-0.1cm}{\cal N}=8\vspace{0.1cm}}_{61} &=&
\sum_{h\in\, {\cal N}=8\, \rm multiplet} 
      M_4^\tree(l_0^{-h}, 6^+, 1^-, -l_1^{h}) \times
       M_6( -l_0^{h}, 2^-, 3^+, 4^+, 5^+, l_1^{-h}) \nn \\ 
	&=&\rho_8\times M_4^\tree(l_0^{-}, 6^+, 1^-, -l_1^{+}) \times
       M_6( -l_0^{+}, 2^-, 3^+, 4^+, 5^+, l_1^{-})\,,
\label{N8GravityCut61}
\end{eqnarray}
where 
\begin{eqnarray}
\rho_8&=& (1 - 8 x + 28 x^2 - 56 x^3 + 70 x^4 - 56 x^5 + 28 x^6
       - 8x^7 + x^8)   \nn \\
    &=&  (1-x)^8 \,,\hskip 2 cm 
x = {\spa{l_1}.1 \spa{l_0}.2 \over \spa{l_0}.1 \spa{l_1}.2} \,.
\label{Rh08DefCut2}
\end{eqnarray}
The relative minus signs between the terms are due to the minus signs
associated with fermionic loops.  Applying the Schouten identity
(\ref{Schouten}) gives,
\begin{equation}
\rho_8 = \left(\frac{\spa{l_1}.{l_0} \spa{1}.2}
{\spa{l_0}.1 \spa{l_1}.2}\right)^8\sim y^{-16}\,t^8\sim (l^{\mu})^{-8}\,.
\label{DoubleCutRhoScaling}
\end{equation}
The scaling of $\rho_8$ makes manifest the cancellations of eight
powers of loop momentum $l$ due to supersymmetry.  In total, the
cut scales as $y^{-8}t^4$ implying the vanishing of the bubble
coefficient which confirms the known result~\cite{OneloopMHVGravity}.
Of the $n-6$ powers of loop momentum that cancel to eliminate the
bubble integrals in the $\NeqEight$ $n$-graviton amplitude, only eight
are due to supersymmetry. {\it Thus, to a large extent, the
one-loop cancellations in the $\NeqEight$ theory originate from cancellations
present in the non-supersymmetric pure gravity case.}  This is a central
result of this paper.

Cases with fewer supersymmetries are similar. For example, the $\NeqOne$
gravity multiplet consisting of a graviton and gravitino in 
the loop, has the corresponding 
factor of,
\begin{equation} 
\rho_1 = 1 -x - x^7 + x^8 \sim y^{-4} t^2 \sim (l^\mu)^{-2}\,,
\label{N1sugra}
\end{equation}
where $x$ is given in \eqn{Rh08DefCut2}. Note that there is an
additional cancellation in this expression, besides the leading one,
reducing the power of loop momentum by two.  Similarly, for an ${\cal
N} = 2$ gravity multiplet, the factor is,
\begin{equation}
\rho_2 = 1 -2 x +x^2 + x^6 - 2x^7 + x^8 \sim y^{-4} t^2 \sim (l^\mu)^{-2} \,,
\label{N2sugra} 
\end{equation} 
so that the additional supersymmetry has not reduced the power count
compared to the $\NeqOne$ case.  We have checked that this pattern
generalizes to higher supersymmetry. Depending on whether the number
of supersymmetries is even or odd we have,
\begin{equation}
\rho_{\mathcal N_{\rm even}}\sim
(l^\mu)^{-{\cal N}_{\rm even}} \,, \hskip 2 cm 
 \rho_{ {\cal N}_{\rm odd}}\sim
(l^\mu)^{-({\cal N}_{\rm odd}+1)} \,.
\label{RhoN}
\end{equation}
It is interesting to note that, according to this pattern, theories
with ${\cal N}\ge 3$ can have no powers of loop momentum in the
numerators of bubble integrals, and therefore no additional rational
terms from here~\cite{Fusing}.\footnote{For ${\cal N}\le 4$ supergravity
there can be rational contributions from box diagrams, because a
reduction of four powers of loop momentum still leaves four powers behind
which under integral reduction gives rational terms. We thank L.
Dixon and C. Boucher-Veronneau for pointing this out.  This is
consistent with the results of ref.~\cite{DunbarNew}.}  A further interesting
bound is ${\cal N}\ge 5$, where the above counting implies that bubble
integrals will not be present in the one-loop
amplitudes~(\ref{BasisIntegrals}).  In addition, there can be at most
three powers of loop momentum in box numerators, which is insufficient
to generate rational terms.

\subsection{A six-point triangle example} 

As a second example, consider the next-to-MHV (NMHV) amplitude
$M_4^\oneloop(1^-, 2^-, 3^-, 4^+,5^+,6^+)$ and the power counting of
the triangle integral shown in \fig{SixBubbleTriangleFigure}(b).  As
described in the previous section, to analyze the power counting in
triangle functions, we consider triple cuts.  Two helicity
configurations of cut legs contribute, 
\begin{eqnarray}
\hspace{-0.7cm}C^{\,\vspace{-0.1cm}{\rm gravity}\vspace{0.1cm}}_{14,25,36}
&=& M_4^{\rm tree}(l_0^-,1^-,4^+,-l_1^+)\times 
    M_4^{\rm tree}(l_1^-,2^-,5^+,-l_2^+)\times 
    M_4^{\rm tree}(l_2^-,3^-,6^+,-l_0^+)\nonumber\\
&&\hspace{-0.3cm}\null
  + M_4^{\rm tree}(l_0^+,1^-,4^+,-l_1^-)\times
    M_4^{\rm tree}(l_1^+,2^-,5^+,-l_2^-)\times 
    M_4^{\rm tree}(l_2^+,3^-,6^+,-l_0^-)\,. \hskip .5 cm 
\label{TripleCutExample}
\end{eqnarray}

First consider the naive power counting of the individual Feynman diagrams
contributing to the tree amplitudes appearing in the cut.  From
\eqn{TripleCutMomentum}, we see that each numerator loop momentum
scales as $t$.  In the denominator, any propagators in the tree
amplitudes $1/(l - K_i)^2$ scale as $1/t$ because of the on-shell
condition $l^2 = 0$.  Using this scaling we thus obtain the leading
behavior from the two vertices and one propagator in each of the
four-point tree-amplitudes giving a total of 
\begin{equation}
(t^2\times t^2/t)^3\sim t^9 \sim (l^\mu)^9\,.
\end{equation}
This corresponds to the eight powers of loop momentum we obtained in
the bubble integrals starting from a hexagon Feynman diagram, because
the bubble integrals require one more step in the reduction compared to 
triangles. 

Now consider power counting in the triangle integral of
\fig{SixBubbleTriangleFigure}(b), while accounting for cancellations
between diagrams.  Using the explicit form of the tree amplitudes and
spinor products collected in (\ref{SpinorsTripleCut}), for the first
term in \eqn{TripleCutExample} we have,
\begin{eqnarray}
 M_4^{\rm tree}(l_0^-,1^-,4^+,-l_1^+)&=&
i\frac{\spb4.1\spa{l_0}.1^7}
       {\spa{l_1}.{l_0}^2\spa1.4\spa4.{l_1}\spa{l_0}.4\spa1.{l_1}}\sim t^2\,.
\end{eqnarray}
The other factors of tree amplitudes in this term are similar, each
also giving a factor of $t^2$.  Thus, the first term in
\eqn{TripleCutExample} scales as $t^6$, indicating six powers of loop
momentum: $(l^\mu)^6$. Similarly, the second term in
\eqn{TripleCutExample} also scales the same way, giving us an overall
scaling of 
\begin{equation}
C^{\,\vspace{-0.1cm}{\rm gravity}\vspace{0.1cm}}_{14,25,36} 
\sim  t^6 \sim (l^\mu)^6 \,.  
\label{TripleCutScaling}
\end{equation}
The
improved scaling is due to cancellations between Feynman diagrams in
the tree amplitudes. 

We may compare this to the results of ref.~\cite{NoTriangleSixPt} for
the ${\cal N}=8$ theory.
As for the bubble integral case, the sum over the 
${\cal N}=8$ multiplet in the loop can be included by a multiplicative factor,
\begin{eqnarray}
C^{\,{\cal N}=8}_{14,25,36}&=& 
\sum_{h\in\, {\cal N}=8\ \rm states}  
  M_4^{\rm tree}(l_0^{-h},1^-,4^+,-l_1^h)\times 
  M_4^{\rm tree}(l_1^{-h},2^-,5^+,-l_2^h)  \\
&& \null \hskip 3 cm \times 
  M_4^{\rm tree}(l_2^{-h},3^-,6^+,-l_0^h)\nonumber\\
&=&\rho_8\times M_4^{\tree}(l_0^-,1^-,4^+,-l_1^+) \times 
  M_4^{\rm tree}(l_1^-,2^-,5^+,-l_2^+)\times 
  M_4^{\rm tree}(l_2^-,3^-,6^+,-l_0^+)\,, \nn
\end{eqnarray}
where,
\begin{eqnarray}
\rho_8    &=&  (1-x)^8 \,,\hskip 2 cm 
x = {\spa{l_1}.1 \spa{l_2}.2 \spa{l_0}.3  \over \spa{l_0}.1 \spa{l_1}.2 
      \spa{l_2}.3}\,.
\end{eqnarray}
From \eqn{SpinorsTripleCut} we have that 
$\spa{l_i}.a \sim t\spa{K^{\flat}_1}.a$, so that the leading $t$ behavior
cancels, giving us, 
\begin{equation}
\rho_8 \sim t^{-8}\,.  
\label{N8TripleCut}
\end{equation} 
Thus, as in the bubble case, the $\NeqEight$ supersymmetry reduces 
the degree of the loop momentum polynomial by eight. 

In total, combining \eqns{TripleCutScaling}{N8TripleCut} shows the
triple-cut integral scales to zero, like $t^{-2}$, for ${\cal N}=8$
supergravity so that the triangle integral is not present in the
amplitude.  However, only because of the cancellations
already present in pure gravity does the triangle integral coefficient
vanish. The
vanishing of this triangle integral coefficient in the $\NeqEight$ theory has
been shown numerically in ref.~\cite{NoTriangleSixPt}.  The
analysis here provides a simple analytic proof of the vanishing.

As for the bubble integral discussed above, the cases with less
supersymmetry are similar. Depending on whether the number of
supersymmetries is odd or even, we have either ${\cal N}+1$ or 
${\cal N}$ powers of canceled loop momentum, respectively.

\subsection{All-$n$ power counting}

We now discuss power counting in the bubble and triangle integrals
for an arbitrary number of external legs.
We will discuss the cancellations in the two
cases, and close with a summary of the checks that we have
performed.

\subsubsection{Bubble integrals}

As described above, in order to perform all-$n$ power counting for
the bubble functions shown in \fig{IntClassesFigure}(c), we determine the
leading power of $y^2/t$ under the large-$y$ scaling properties of the
tree amplitudes contributing to the two-particle cuts of
\fig{CutsFigure}(c).
There are two possible  helicity configurations
crossing the two-particle cuts.  Either the legs crossing
the cuts are of the same helicity or they are of opposite helicity.
If they are of opposite helicity, then the contribution to the
cuts is,
\begin{eqnarray}
C_2^{(-,+)} &=&
M_{m+2}^\tree(l_0^-,  i_1, \ldots i_{m},   -l_1^+ ) \times
M_{n-m+2}^\tree(l_1^-\,, i_1^{'}, \ldots i_{n-m}^{'}, -l_0^+) \nn \\
&& \null 
+ M_{m+2}^\tree(l_0^+, i_1, \ldots i_{m}, -l_1^- ) \times
M_{n-m+2}^\tree(l_1^+\,, i_{1}^{'}, \ldots i_{n-m}^{'}, -l_0^-)\,, \hskip 1 cm 
\label{TwoParticleMPCut}
\end{eqnarray}
where we identify $l_0$ with the loop momentum $l$.
If the two legs crossing the cut are of the same helicity
the contribution is, 
\begin{eqnarray}
C_2^{(+,+)} &=& 
M_{m+2}^\tree(l_0^+, i_1, \ldots i_{m}, -l_1^+ ) \times
M_{n-m+2}^\tree(l_1^-\,, i_{1}^{'}, \ldots i_{n-m}^{'}, -l_0^-) \nn\\
&& \null
+
M_{m+2}^\tree(l_0^-, i_1, \ldots i_{m}, -l_1^- ) \times
M_{n-m+2}^\tree(l_1^+\,, i_{1}^{'}, \ldots i_{n-m}^{'}, -l_0^+) \,. 
\label{TwoParticlePPCut}
\end{eqnarray}
%
%%%%%%%%%%%%%%% TABLE %%%%%%%%%%%%%%%%%%%%%%%%%%%%%%%%%%%
\def\hs{\null \hskip .2 cm \null }
\begin{table*}[t]
{\vskip .4 cm}
\begin{tabular}{|l|c|c|c|c|}
\hline\hline
\hs $ (h_{l_0},-h_{l_1})$\hs & \hs$(-,+)$ \hs &\hs$(+,-)$\hs  &\hs$(+,+)$ \hs &\hs $(-,-)$ \hs  \\
\hline
\hs $y$-scaling of $M^\tree_n$ \hs&$y^4t^{2}$  &$y^4t^{-6}$&$y^{-4}t^{2}$ &$y^{-4}t^{2}$  \\
\hline\hline
\end{tabular}
\vspace{0.4cm}
\caption{The large $y$ scaling of tree-amplitudes appearing in the
two-particle cuts of a one-loop amplitude.  The scaling depends on the
helicities of the two legs carrying loop momentum, indicated in the first row. 
The amplitudes are evaluated using the cut kinematics  
in \eqns{BubbleCut4Vector}{DoubleCutMomDef}.
\label{TwoParticleCutTable} }
\end{table*}
%%%%%%%%%%%%%%%%%%%%%%%%%%%%%%%%%
%
For cases where the tree amplitudes in the cuts are either MHV or
\MHVbar, we can read off the scaling in $y$, by choosing leg $l_0$ and
$l_1$ to correspond to legs $1$ and $n$ in
\eqns{ThreeFourPointTree}{BGKMHV}, or their parity conjugates.  With
this choice, the scaling properties in $y$ and $t$ from
\eqn{DoubleCutMomDef} are manifest in each term.  Choosing the
negative helicity legs in $a$ and $b$, to match the tree amplitude
appearing in the cuts, we can easily evaluate all cases, collected in
\tab{TwoParticleCutTable}.  For more general helicities we do not have
a complete proof, but we have numerically checked that the scaling in
\tab{TwoParticleCutTable} is correct for {\it all} helicity
configurations up to ten points.  We have also constructed an
$n$-point proof for the cases where opposite helicities cross the cut,
outlined in \sect{CancellationSection}.

Using \tab{TwoParticleCutTable}, we
can then read off the maximum scaling of the cuts (\ref{TwoParticleMPCut})
and (\ref{TwoParticlePPCut}) at large $y$,
\begin{eqnarray}
C_2^{(-,+)} &\sim& y^4t^{2} \times y^4t^{-6} \sim (l^\mu)^4 \,, \nn \\
C_2^{(+,+)} &\sim& y^{-4}t^{2} \times y^{-4}t^{2} \sim (l^\mu)^{-4} \,. 
\label{BubbleScaling}
\end{eqnarray}
Thus, for the case where opposite helicities cross the cut, up to four
powers of loop momentum appear in the numerator of the bubble
integral.  For the case where like helicities cross the cut, the
bubble contributions cancel completely, as indicated by the negative
power of $y$.  Both cases exhibit non-trivial cancellations compared
to individual Feynman diagrams. For $n \le 10$ external gravitons our
numerical analysis directly confirms \eqn{BubbleScaling} for all helicity
configurations.

We may compare these results to the corresponding ones for an ${\cal
N}$-extended supergravity.  The case with identical helicities
crossing the cut is the same in the two theories and there are no
contributing bubble integrals.  For the case of opposite helicities crossing
the cut, we need to sum over the contributions of the entire
super-multiplet.  For the case of MHV or \MHVbar\ amplitudes, it is
simple to carry out the sum over the multiplet, using the
supersymmetry Ward identify (\ref{SWIforMHV}).  As for the six-point
case, with opposite helicity legs crossing the cuts, 
the net effect from the super-multiplet sum is the additional overall
factor of $\rho_{\cal N}$ given in \eqn{RhoN}.

Specializing to ${\cal N}=8$, the
bubble contributions all scale as $y^{-4}t^2$, so there are no bubble
integrals present in the MHV amplitudes, as already discussed in
refs.~\cite{OneloopMHVGravity,NoTriangleSixPt}. As for the six-point
example above, we find that the supersymmetry only reduces the number 
of loop momenta by eight powers when opposite helicities
cross the cut, with the remaining cancellations
present even in pure gravity. When like helicities cross the cut,
the $\NeqEight$ cancellations are identical to those of pure gravity.

\subsubsection{Triangle integrals}

%%%%%%%%%%% TABLE %%%%%%%%%%%%%
\begin{table*}[t]
%\vskip .4 cm
\begin{tabular}{|l|c|c|c|c|}
\hline\hline
\hs $(h_{l_i},-h_{l_{i+1}})$ \hs &\hs $(-,+)$\hs &\hs$(+,-)$ \hs&
                              \hs $(+,+)$\hs &\hs $(-,-)$ \hs\\
\hline 
\hs $t$-scaling of $M_{n}^{\tree}$ \hs & $t^{2}$    & $t^{2}$  & 
                                        $t^{-6}$   & $t^{2}$ \\
\hline\hline
\end{tabular}
\vspace{0.4cm}
\caption{The leading $t$-scaling of tree amplitudes appearing in the
triple-cuts. The tree amplitudes are evaluated using the spinor inner
products in \eqn{SpinorsTripleCut}.  The first row corresponds to
the helicities of the cut legs carrying loop momentum.
\label{TripleCutTable} }
\end{table*}  
%%%%%%%%%%%%%%%%%%%%%%%%%%%%%%%%

The power counting of the triangle integral of
\fig{IntClassesFigure}(b) is determined by the large-$t$ scaling of
the three tree amplitudes composing the triple cut of
\fig{CutsFigure}(b).
For example, the contribution, where each tree amplitude has opposite helicity
cut legs, is given by,
{\small
\begin{eqnarray}
C_3&=& 
M_{_{m+2}}^\tree\!(l_0^+\,,i_{_1},\ldots i_{_m}, -l_1^- ) \times
M_{_{m'+2}}^\tree\!(l_1^+\,,i^{\,'}_{_1},\ldots i^{\,'}_{_{m'}}, -l_2^-)\times
M_{_{m''+2}}^\tree\!(l_2^+\,,i^{\,''}_{_1}, \ldots i^{\,''}_{_{m''}}, -l_0^-) \nn\\
&&\hspace{-.5cm}
\null 
+M_{_{m+2}}^\tree\!(l_0^-\,,i_{_1},\ldots i_{_m}, -l_1^+ ) \times
M_{_{m'+2}}^\tree\!(l_1^-\,,i^{\,'}_{_1},\ldots i^{\,'}_{_{m'}}, -l_2^+)\times
M_{_{m''+2}}^\tree\!(l_2^-\,,i^{\,''}_{_1}, \ldots i^{\,''}_{_{m''}}, -l_0^+)\,,
\hskip 1.5 cm 
\end{eqnarray}
}
where $m+m'+m''=n$ for an $n$-point amplitude.

Again,  with MHV or \MHVbar\ tree amplitudes appearing in the triple cuts,
it is a simple matter to obtain the scaling pattern in
\tab{TripleCutTable} from \eqns{ThreeFourPointTree}{BGKMHV} and their
parity conjugates, by
choosing the legs carrying loop momentum to be legs $1$ and $n$.
For {\it all} helicity configurations, we have numerically
checked that the table is correct up to ten points, using the
kinematics of \eqns{TripleCutMomentum}{SpinorsTripleCut}.
An all-$n$ proof for the particular helicity configurations $(\pm,\mp)$ 
crossing the cut will be outlined in \sect{CancellationSection}.

Given the scalings in \tab{TripleCutTable} we see that no product of
the three tree amplitudes in a triple cut can have worse scaling than,
\begin{equation}
C_3 \sim t^2\times t^2\times t^2 \sim (l^\mu)^6\,,
\end{equation}
including contributions from integral reductions from higher-point
integrals.  Again, this is significantly better than the behavior
obtained from an $n$-gon Feynman diagram which can generate triangles
integrals with up to $n+3$ powers of loop momentum in their
numerators.

For the cases where the cuts are composed of purely MHV and \MHVbar\
amplitudes, we can easily compare to the results of $\NeqEight$
supergravity, by making use of the supersymmetry
identity~(\ref{SWIforMHV}).  In all cases, we obtain an overall
scaling of $t^{-2} \sim (l^\mu)^{-2}$, including any additional
cancellations from supersymmetry, which means that triangle integrals
are not present.  As in our previous examples, the key $n$-dependent
cancellations are already present in the non-supersymmetric case.

\subsubsection{Summary of checks}
In summary our checks confirm cancellations
for pure Einstein gravity within each triangle and bubble contribution.
We confirm these cancellations
when the tree amplitudes appearing the cuts determining the
integrals are from among the following:
\begin{enumerate}
\item An $m$-point MHV amplitude.
\item An $m$-point \MHVbar\ amplitude.
\item Any helicity configuration with up to ten gravitons.
\item An $m$-point amplitude with loop-momentum helicities
$(\pm,\mp)$.\footnote{This check will be discussed below in
\sect{CancellationSection}.}
\end{enumerate}
This demonstrates that the one-loop cancellations hold through ten
points in pure gravity for all helicity configurations and argues
strongly that it continues to hold beyond this.  One might be
concerned that the results for the MHV and \MHVbar\ amplitudes were
derived using the BGK formula, which has been tested only up to 11
points and not beyond this. This is not a real restriction here, as
the information we extract from the BGK formula is its scaling
properties, which will be shown to hold for all $m$ by an independent
line of reasoning in \sect{CancellationSection}.

For cases with only MHV or \MHVbar\ amplitudes in the cuts we analyzed
the additional cancellations in ${\cal N}$-extended supergravity
theories by summing over the super-multiplet.  When opposite
helicities cross the cut, depending on whether ${\cal N}$ is even or
odd we have additional cancellations of ${\cal N}$ or $({\cal N}+1)$
powers of loop momentum, respectively.  In particular, for $\NeqEight$
supergravity, the no-triangle hypothesis follows from a combination of
cancellations, with at most eight powers of loop momentum in the
numerator canceled by supersymmetry and the remaining cancellations
already present in pure gravity.

%%%%%%%%%%%%%%%%%%
\section{Relations between different cancellations}
\label{CancellationSection}

We now discuss a proof
that the one-loop cancellations described above are
simply related to the tree-level large-$z$ cancellations under the shift
(\ref{SpinorShift}). We also present a heuristic argument using
factorization to explain the pure gravity cancellations, similar to
the one used to propose the ``no-triangle hypothesis'' for $\NeqEight$
supergravity in ref.~\cite{NoTriangle}.

\subsection{Relations between tree and triangle-bubble scaling properties}

Consider a generic tree amplitude,
\begin{eqnarray}
M_n^\tree(l_0^{h_0},-l_1^{h_1},i_1,\ldots ,i_{n-2})\,,
\label{generictree}
\end{eqnarray}
where $h_0$ and $h_1$ denote the helicities --- for gravitons these
take on the values of $\pm\,2$.  We now observe some simple relations
between the different scalings.  Suppose that after a
$\Shift{l_0}{-l_1}$ shift (\ref{ShiftMomentum}), and after accounting
for all cancellations, we find that the tree amplitude scales in the
large-$z$ limit as,
\begin{eqnarray}
\hspace{0.1cm}\Shift{l_0}{-l_1}:\quad\quad  M_n^\tree(z)\,\sim\, z^{m}\,,
\label{InitialShift}
\end{eqnarray}
where the value of $m$ depends upon the helicities. Using
\tab{ShiftZTable}, we see that the scaling of the flipped
$\Shift{-l_1}{l_0}$ shift, at large $z$, is related to the above by,
\begin{equation}
\Shift{-l_1}{l_0}:\quad   M_n^\tree(z)\sim z^{m+2(h_1-h_0)}\,.
\label{FlippedShift}
\end{equation}
Similarly, consulting tables~(\ref{ShiftZTable}) and
(\ref{TwoParticleCutTable}), we find that the large-$y$ scaling of the
two-particle cut momentum parameterization~(\ref{BubbleCut4Vector}) is
given by,
\begin{eqnarray}
\hspace{2.1cm}\mbox{$y$-scaling:}\quad\quad M_n^\tree(y,t)
  \sim \big(y^2/t\big)^{m}\,y^{2(h_1-h_0)}\,.
\label{yScaling}
\end{eqnarray}
From tables~(\ref{ShiftZTable}) and (\ref{TripleCutTable}),
we observe 
that under the loop-momentum parametrization for the
triple cut~(\ref{TripleCutMomentum}),
\begin{eqnarray}
\hspace{-0.4cm}\mbox{$t$-scaling:}
\quad\quad M_n^\tree(t)\,\sim\, t^{m}\,t^{-2h_0}\,.
\label{tScaling}
\end{eqnarray}
This indicates that all the scalings are related up to universal
helicity dependent factors.  Were these observed relations to hold on
very general grounds, the non-trivial cancellations in the Feynman
diagrams in one of the scalings would necessarily imply non-trivial
cancellations in the other scalings. In fact, this turns out to be the
case, as we now argue.

To establish the above scaling relations, we write the scattering
amplitude (\ref{generictree}) in an abstract form making the helicity
weights manifest.  The tree-level scattering amplitudes are rational
functions in all spinor variables.  Using momentum conservation,
$l_1=l+k_{i_1}+\ldots +k_{i_{n-2}}$, we trade momentum $l_1$ for
$l_0$, absorbing the spinor weight of the leg carrying momentum $l_1$
into factors of $\spa{l_0}.{l_1}$.  This allows us to write an
amplitude with the legs carrying the momentum $l_0$ and $l_1$ with the
corresponding helicities $h_0$ and $h_1$ as
\begin{eqnarray}
M_n^\tree(l_0^{h_0},-l_1^{h_1},i_1,\ldots, i_{n-2}) =
\spa{l_0}.{l_1}^{-2h_1}\, {N_{s'}(\lambda, \tilde \lambda) \over 
  D_{s''}(\lambda, \tilde \lambda)} \,,
\label{generictreeform}
\end{eqnarray}
where $N_{s'}(\lambda, \tilde \lambda)$ and $D_{s''}(\lambda, \tilde
\lambda)$ are polynomials in spinors $\lambda$ and $\tilde \lambda$,
satisfying $(l_0)_{a \dot a} = \lambda_a \tilde \lambda_{\dot a}$
and we have suppressed the dependence on external momenta. The
polynomials $N_{s'}(\lambda, \tilde \lambda)$ and $D_{s''}(\lambda,
\tilde \lambda)$ have definite helicity weight --- the number of
$\lambda$s minus the number of $\tilde \lambda$s --- denoted by
subscripts $s'$ and $s''$.  For the total amplitude to have proper
helicity weight with respect to the $\lambda$ and $\tilde\lambda$, the
weights $s'$ and $s''$ must be related to the helicities by,
\begin{eqnarray}
 s'-s''=2(h_1-h_0)\,.
\end{eqnarray}
In the scaling limits considered above, the polynomials are dominated
by the highest degree monomials in the $\lambda$ and $\tilde \lambda$
variables,
\begin{eqnarray}
N_{s'}(\lambda, \tilde \lambda)&\sim &
\tilde\lambda_{\dot a_1}\cdots\tilde\lambda_{\dot a_{r'}}\,\lambda_{b_1}\cdots\,\lambda_{b_{r'+s'}}\,N^{\dot a_1
\ldots \dot a_{r'}\,b_{1} \ldots b_{r'+s'}}\,, \nn\\
D_{s''}(\lambda, \tilde \lambda) &\sim &
\tilde\lambda_{\dot a_1}\cdots\,\tilde\lambda_{\dot a_{r''}}\,
\lambda_{b_1}\cdots\,\lambda_{b_{{r''+s''}}}\,
D^{\dot a_1 \ldots \dot a_{r''}\,b_{1}\ldots b_{{r''+s''}}}\,.
\label{monomials}
\end{eqnarray}
A priori, the values of $r'$ and $r''$ can be functions of the number
of external legs and their helicities.

This general form of the amplitude may now be probed in the various
scaling limits.   For the
shift $\Shift{l_0}{-l_1}$ of the amplitude \eqn{generictree} we have
the scaling,
\begin{eqnarray}
\Shift{l_0}{-l_1}:\quad   M_n^\tree(z)\sim z^{r'-r''}\equiv z^{m},
\end{eqnarray}
where we introduced the variable $m={r'-r''}$ to match
\eqn{InitialShift}.  Applying the $t$ and $y$ scalings to
\eqns{generictreeform}{monomials} immediately gives us
\eqns{tScaling}{yScaling}.  Similarly we obtain the flipped
$z$-shift result (\ref{FlippedShift}). Notice that the value 
of $m$ does not follow from this analysis and has to be derived 
by other means.

The relations between the various scalings thus hold on very general
grounds and are applicable to a wide variety of cases including
amplitudes with general matter content.  In our analysis we used
that the monomials in \eqn{monomials} do not vanish in the scaling
limits. As will be discussed elsewhere, this always holds for the
scaling limits discussed here.

The $z$-scaling for the shift $\Shift{-}{+}$ in~\tab{ShiftZTable} has
been proven rigorously for all $n$ in
ref.~\cite{CachazoLargez}. Direct comparison with this then allows for
an easy proof of the corresponding $(-,+)$ case of the $y$ and $t$
scaling. Because the shift $\Shift{+}{-}$ is the flipped version of
the shift $\Shift{-}{+}$, the two are related by the constraints of
spinor weight. This then gives a proof of the large-$z$ scaling of the
shift $\Shift{+}{-}$ in~\tab{ShiftZTable} for all $n$. In turn this
can be used to prove the $(+,-)$ cases in \tab{TwoParticleCutTable}
and \tab{TripleCutTable}, as well.  The $\Shift{\pm}{\pm}$ and
$(\pm,\pm)$ cases, however, remain unproven beyond ten points for
non-MHV amplitudes.  For MHV amplitudes, the supersymmetry identity
(\ref{SWIforMHV}) allows us to apply the $\Shift{-}{+}$ proof to these
cases as well~\cite{CachazoLargez}.

These results demonstrate that the bubble-triangle cancellations
occur whenever all tree amplitudes composing a cut have loop-momentum
helicity configurations $(\pm,\mp)$.  It also shows that if the
$\Shift{+}{+}$ and $\Shift{-}{-}$ entries in \tab{ShiftZTable} are
valid for any number of legs, the bubble-triangle cancellations
also hold for any number of legs for the other helicity configurations
of the cut legs.

\subsection{Heuristic relation between cancellations and 
factorization properties}

In addition, the bubble-triangle cancellations may be understood heuristically
as a consequence of the stringent factorization properties
of one-loop amplitudes.  As any intermediate momentum $K^\mu \equiv
k_i^\mu+\ldots +k_{i+r+1}^\mu$ goes on-shell ($K^2 \rightarrow 0$), a
one-loop amplitude factorizes into lower-point amplitudes (along with
a universal ``factorization function'' for infrared singular
amplitudes~\cite{BernChalmers}).  As we demonstrated through ten
points, for all helicities, no graviton amplitudes can have tensor
triangle or bubble integrals with more than six or four powers of loop
momentum in the numerators, respectively.  Therefore, in {\it all}
factorization limits of higher-point amplitudes we cannot encounter
these integral functions or the results of reducing them to scalar
integrals.  Moreover, the same type of argument holds for
factorizations where two of the external momenta become
collinear\footnote{Although there is no kinematic pole for collinear
limits with real momenta, there is a universal phase
singularity~\cite{OneloopMHVGravity} or equivalently universal
behavior under factorization with complex momenta.} or where one of
the external momenta becomes soft.  Although this does not constitute
a proof\footnote{In principle, functions can be present which have no
poles in any channel. An example of such a function which may occur in
abelian gauge theories, may be found in eq.~(14) of
ref.~\cite{MinneapolisReview}.} that the unwanted tensor triangle and
bubble integrals cannot appear, we know of no counterexample, in
either gravity or non-abelian gauge theory where this type of
factorization bootstrap argument has failed to produce the correct
result at one loop.

Indeed, this argument is similar to the one used in ref.~\cite{NoTriangle} to
propose the no-triangle hypothesis for the $\NeqEight$ 
theory. In that case, there were no bubble or triangle
integrals at all at lower points, predicting that bubble or triangle
integrals should not appear at higher points as well.

%%%%%%%%%%%%%%%%%%%%%%%%

\section{Conclusions}
\label{ConclusionSection}

In this paper we studied cancellations in pure gravity one-loop
amplitudes, pointing to the existence of novel ultraviolet
cancellations in generic gravity theories at higher loops.  At one
loop, without the novel cancellations, the two-derivative coupling of
gravity would imply that under integral reductions of $n$-point
amplitudes we naively would obtain bubble integrals with up
to $n+2$ powers of loop momentum in their numerators.  Similarly, we
would obtain triangle integrals with up to $n+3$ powers of loop
momentum in their numerators.  Instead, under more careful scrutiny, we
found that the triangle and bubble integrals resulting from integral
reductions have no more than six or four powers of loop momentum,
respectively.

By comparing the pure gravity case to ${\cal N}$-extended supergravity
in various examples, we disentangled the supersymmetric cancellations
in these theories from the ones which are generic to gravity.
Assuming the universality of the gravity cancellations,
one-loop ${\cal N} {\ge5}\,$-extended supergravity amplitudes should be
cut-constructible~\cite{Fusing} using only four-dimensional momenta in
the cuts. That is, they are completely determined by their absorptive
parts.  Similarly, the ``no-triangle hypothesis'' of the $\NeqEight$
theory~\cite{NoTriangle,NoTriangleSixPt} may be thought of as a
consequence of the combination of the pure gravity cancellations 
and supersymmetric cancellations.  It is interesting to note that
conventional superspace power counting is sensitive to only the latter
types of cancellations.

The unitarity method, together with a new spinor-based integration
method~\cite{Forde}, allows us to link directly cancellations
occurring in one-loop amplitudes to cancellations in tree-level
amplitudes. This follows an early version of scaling arguments,
used to imply that bubble integrals should not appear in $\NeqEight$
supergravity amplitudes~\cite{NoTriangleSixPt}.  Our approach allows
us to determine straightforwardly the maximum powers of loop momentum
that appear in the triangle and bubble integrals, including all feed
downs from integral reductions.  Although we have not constructed a
complete proof of one-loop cancellations for all possible gravity
amplitudes, we numerically demonstrated their existence up to ten
points for {\it all} pure gravity helicity amplitudes, and
analytically for special helicity configurations for all $n$.  We have
also outlined a proof for all contributions to triangle and bubble
integrals whose cuts are composed of tree amplitudes with opposite
graviton helicities on the legs carrying loop momentum.

The cancellations discussed here are connected to recently identified
properties of gravity tree amplitudes.  In the usual Lagrangian
formulation an infinite set of vertices are needed to construct the
scattering amplitudes.  This may be contrasted with on-shell
recursion relations~\cite{GravityRecursion}
which remarkably construct all gravity tree amplitudes in
$D=4$, starting {\it only} from three vertices.  This property 
has been recently proposed as a means of classifying
theories~\cite{CachazoConstuctible}.  For the recursion relations to
hold cancellations under certain large complex deformations of the 
amplitudes are necessary.  The existence of these cancellations are
rather obscure in Feynman diagrams. A proof of the
cancellations required to have valid on-shell recursion relations
has recently been given in ref.~\cite{CachazoLargez}.
Here we showed that related cancellations exist at one loop.

%%%%%%%%% FIGURE %%%%%%%%%%%%%%%
\begin{figure}[t]
\centerline{\epsfxsize 3.1 truein \epsfbox{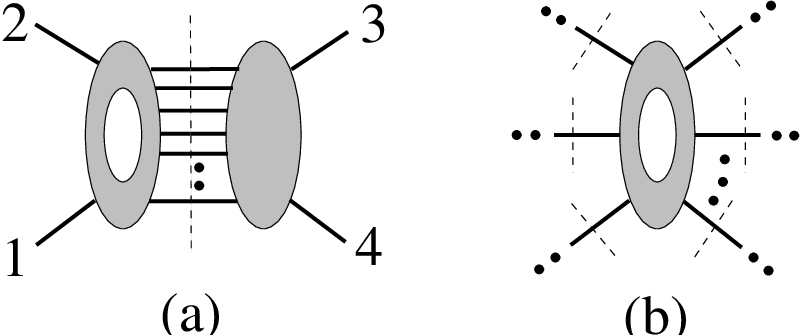}}
\caption[a]{\small From the unitarity cuts, cancellations in
one-loop subamplitudes imply higher-loop
cancellations.  The cut (a) is an $L$-particle cut of an $L$-loop
amplitude, isolating a one-loop amplitude on the left side of the
cut. Diagram (b) denotes a generalized cut that isolates a particular
one-loop subamplitude. If a leg is external to the entire amplitude,
it should not be cut.  In ref.~\cite{Finite}, these cuts were used to 
argue that in $\NeqEight$ supergravity the one-loop no-triangle hypothesis
implies the existence of non-trivial cancellation at higher-loop orders.}
\label{NCutFigure}
\end{figure}
%%%%%%%%%%%%%%%%%%%%%%%%%%%%%%%%

Unitarity implies lower-loop cancellations necessarily induce
higher-loop cancellations.  In particular, as shown in \fig{NCutFigure}(a), the
cancellations in the $(L+2)$-point one-loop amplitude appearing in the
$L$-particle cut of the $L$-loop amplitude, necessarily imply the
existence of cancellations in the $L$-loop amplitude.  Moreover, as
indicated in \fig{NCutFigure}(b) any one-loop sub-amplitude isolated
by cuts necessarily must have the cancellations found in our one-loop
investigations.  In ref.~\cite{Finite}, these cuts, together with the
one-loop no-triangle hypothesis, were used to suggest that $\NeqEight$
supergravity may be ultraviolet finite.  These cancellations, as well
as others not implied by the no-triangle hypothesis, were then
confirmed by explicit calculation at three
loops~\cite{GravityThree}. Similarly, the results of this paper
suggest that novel cancellations at higher loops should exist even in
non-supersymmetric theories, though the cancellations will not be as
strong as for $\NeqEight$ supergravity.  

In order to check this, it would be rather useful to carry out
explicit higher-loop studies of gravity theories. Of course,
in pure gravity a two-loop divergence does
exist~\cite{GoroffSagnotti}, but an open question is to determine the
critical dimension in which the divergences first appear, with and
without matter, as the loop order increases.  It would also be helpful
to translate any higher-loop cancellations into the language of
effective actions. For example, for the case of $\NeqEight$
supergravity, at three loops, 14 powers of external momenta can be
extracted from the numerators of all loop momentum integrals, giving a
contribution to the effective action of the generic form ${\cal D}^6
R^4$ multiplied by integrals, instead of the generic form ${\cal D}^4
R^4$, which would have been obtained if additional cancellations had
not been present~\cite{GravityThree}. Here $R^4$ denotes an
$\NeqEight$ supersymmetric contraction of Riemann
tensors~\cite{SuperGravity}, and ${\cal D}$ denotes a generic
space-time covariant derivative, with Lorentz indices contracted
appropriately.

An important calculation that will help to answer the question of whether
$\NeqEight$ supergravity is finite, and to further constrain possible
superspace explanations, is to evaluate the four-point four-loop
$\NeqEight$ amplitude. This would tell us whether it has the same
power counting as $\NeqFour$ super-Yang-Mills theory, as already
established at one~\cite{GSB, OneloopMHVGravity, NoTriangle,
NoTriangleSixPt,TravagliniN8OneLoop}, two~\cite{BDDPR} and three
loops~\cite{GravityThree}.  The four-loop calculation should be
feasible, though non-trivial, following the same methods as used at
three loops in ref.~\cite{GravityThree}. If $\NeqEight$ supergravity
can be shown to be finite, an obvious question is whether there are
other such theories.  The absence of bubble integrals at one loop
hints that theories with ${\cal N} \ge 5$ supersymmetries are candidate
ultraviolet finite theories.  One may wonder whether finiteness in
gravity theories is connected to the possible existence of topological
string theory descriptions~\cite{WittenTopologicalString}.  Recent
developments in constructing such a string for gravity may be found in
ref.~\cite{N8TwistorString}.

There are a number of other issues that would be interesting to
explore.  It would be important to complete a proof that any one-loop
amplitude in any theory based on the Einstein-Hilbert action coupled
to matter has no more than four powers of loop momentum in the
numerators of bubble integrals, and no more than six powers in
triangle integrals, including those generated via integral reductions.
To simplify the analysis, we evaluated the cuts in $D=4$ where we have
powerful spinor methods for generating amplitudes and for evaluating
the resulting integrals.  It would be useful to instead carry out a
power counting analysis in $D=4-2\eps$ dimensions to properly account
for dimensional regularization.  It would also be helpful to have a
general one-loop analysis of the relative power counting behavior
between supersymmetric and non-supersymmetric amplitudes.  Finally, it
would be important to re-express the cancellations in terms of
effective actions.

In summary, the results of this paper point to the existence of novel
loop-level cancellations in generic point-like theories of quantum
gravity based on the Einstein-Hilbert action.  As we showed, these
cancellation are related to previously identified tree-level
cancellations in gravity amplitudes under large complex
deformations~\cite{GravityRecursion,CachazoLargez,CachazoConstuctible}.
The cancellations suggest that, in general, the
perturbative ultraviolet properties of quantum gravity theories may be
tamer than anticipated, pointing to an improved ultraviolet behavior
in the Wilsonian effective action as the cutoff is varied, independent
of supersymmetry.  As already suggested in refs.~\cite{Finite,
GravityThree} for $\NeqEight$ supergravity, the combination of these
cancellations with supersymmetric ones may be sufficient to render the
theory ultraviolet finite to all loop orders.  Further work will be
required to confirm these proposals.

\section*{Acknowledgements}
\vskip -.3 cm 
We especially thank David Kosower for helpful comments
and suggestions.  We also thank Nathan Berkovits, Gordon Chalmers,
Lance Dixon, Dave Dunbar, Renata Kallosh, Prem Kumar, Asad Naqvi, 
Carlos N\'u\~nez, Radu Roiban and Edward Witten for helpful comments 
and discussions.  We thank Academic Technology Services at UCLA for 
computer support.  This research was supported by the US Department 
of Energy under contracts DE--FG03--91ER40662 and DE--AC02--76SF00515.

%%%%%%%%%%%%%%%%%%%%%%%%%%%%%%%%%%%%%%%%%%%%%%%%%%%%%%%%%%

\end{document}